\documentclass[11pt]{article}
\usepackage{amssymb,amsmath}
\usepackage{bm} 
\usepackage{booktabs} 
\usepackage{array}
\usepackage{latexsym}
\usepackage{graphicx}
\usepackage{color}
\usepackage{datetime}
\usepackage[nosort]{cite}
\usepackage{verbatim}
\usepackage{enumerate}
\usepackage{chngpage} 
\usepackage{mathrsfs}
\usepackage{euscript}
\usepackage{psfrag}

\usepackage{datetime}

\usepackage[nosort]{cite}
\usepackage{chngpage} 
\usepackage{setspace}
\usepackage{tensor}

\usepackage{tensor}
\newcommand{\ind}[1]{\indices{#1}}

\usepackage{mciteplus}

\usepackage[colorlinks=true,      linkcolor=blue,      urlcolor=blue,      
            filecolor=blue,      citecolor=blue,       pdfstartview=FitH,     
						pdfpagemode=UseNone,      bookmarksopen=true]{hyperref}  
\usepackage[all]{hypcap}     

\newcommand{\comm}[1]{} 

\def\({\left(}
\def\){\right)}
\def\[{\left[}
\def\]{\right]}

\def\coeff#1#2{{\textstyle \frac{#1}{#2}}}

\def\One{{\hbox{ 1\kern-.8mm l}}}

\def\barray{\begin{array}}
\def\earray{\end{array}}
\def\be{\begin{equation}}
\def\ee{\end{equation}}
\def\bea{\begin{eqnarray}}
\def\eea{\end{eqnarray}}
\def\bal{\begin{align}}
\def\eal{\end{align}}

\def\hI{{\hat{I}}}
\def\hJ{{\hat{J}}}
\def\hmu{{\hat{\mu}}}
\def\hnu{{\hat{\nu}}}
\def\hrho{{\hat{\rho}}}

\newcommand{\abs}[1]{\ensuremath{\left|#1\right|}}

\numberwithin{equation}{section} 

\definecolor{cardinal}{rgb}{0.6,0,0}
\definecolor{darkgreen}{rgb}{0,0.4,0}
\definecolor{golden}{rgb}{0.92, 0.7, 0}
\definecolor{midnight}{rgb}{0, 0, 0.5}
\definecolor{darkblue}{rgb}{0, 0, 0.7}
\definecolor{purple}{rgb}{0.5, 0, 0.5}

\def\oneone{\rlap 1\mkern4mu{\rm l}}
\def\IC{\mathbb{C}}
\def\Neql#1{{\cal N}\!=\!{#1}}

\def\IP{\mathbb{P}}
\def\IR{\mathbb{R}}

\def\IT{\mathbb{T}}

\def\cB{{\cal B}}

\def\cD{{\cal D}}
\def\cF{{\cal F}}

\def\cK{{\cal K}}
\def\cL{{\cal L}}
\def\cM{{\cal M}}
\def\cN{{\cal N}}

\def\cR{{\cal R}}

\def\cX{{\cal X}}

\def\nBPS#1{$\frac{1}{#1}$-BPS}

\begin{document}


\begin{flushright}
IPHT-T20/037
\end{flushright}

\phantom{AAA}
\vspace{-10mm}

\begin{flushright}
%
%
\end{flushright}

\vspace{1.9cm}

\begin{center}

{\LARGE \bf Microstate Geometries from Gauged Supergravity} \\
{\LARGE \bf in Three Dimensions}

{\huge {\bf  }}

\vspace{1cm}

{\large{\bf {Daniel R. Mayerson$^1$,~Robert A. Walker$^{1,2}$   and  Nicholas P. Warner$^{1,2,3}$}}}

\vspace{1cm}

$^1$Universit\'e Paris Saclay, CNRS, CEA,\\
Institut de Physique Th\'eorique,\\
91191, Gif sur Yvette, France \\[12 pt]
\centerline{$^2$Department of Physics and Astronomy}
\centerline{and $^3$Department of Mathematics,}
\centerline{University of Southern California,} 
\centerline{Los Angeles, CA 90089, USA}

\vspace{10mm} 
{\footnotesize\upshape\ttfamily daniel.mayerson @ ipht.fr, walkerra @ usc.edu, warner @ usc.edu} \\

\vspace{2.2cm}
 
\textsc{Abstract}

\end{center}

\begin{adjustwidth}{3mm}{3mm} 
 
\vspace{-1.2mm}
\noindent
The most detailed constructions of microstate geometries, and particularly of superstrata, are done using $\mathcal{N} = (1,0)$ supergravity coupled to two anti-self-dual tensor multiplets in six dimensions.   We show that  an important sub-sector of this theory has a consistent truncation to a particular gauged supergravity in three dimensions.  Our consistent truncation  is closely related to those recently laid out by Samtleben and Sar{\i}o\u{g}lu \cite{Samtleben:2019zrh}, which enables us to develop complete uplift formulae from the three-dimensional theory to six dimensions. We also find a new family of multi-mode superstrata, indexed by two arbitrary holomorphic functions of one complex variable, that live within our consistent truncation and use this family to provide extensive tests of our consistent truncation.  We discuss some of the future applications of having an intrinsically three-dimensional formulation of a significant class of microstate geometries.

\end{adjustwidth}

\thispagestyle{empty}
\newpage


\baselineskip=17pt
\parskip=5pt

\setcounter{tocdepth}{2}
\tableofcontents

\baselineskip=15pt
\parskip=3pt

\section{Introduction}
\label{sec:Intro}

 
The construction of BPS/supersymmetric microstate geometries in five and six dimensions is now a well-developed art \cite{Bena:2007kg,Bena:2013dka,Bena:2015bea,Bena:2016ypk,Bena:2017xbt,Heidmann:2019xrd,Warner:2019jll}.  In particular, superstrata represent one of the broadest families of such geometries and have the advantage of a highly developed holographic dictionary \cite{Kanitscheider:2006zf,Kanitscheider:2007wq,Giusto:2004id,Ford:2006yb, Lunin:2012gp, Giusto:2013bda,Bena:2015bea,Bena:2016ypk,Bena:2017xbt,Shigemori:2020yuo}.  Superstrata are based on the D1-D5 system, whose underlying CFT is created by open strings stretched between the branes, and so the field theory has a world-volume along the common directions of the branes.  The most general known families of superstrata are supersymmetric and  encode a variety of left-moving excitations of the CFT.  Encoding such momentum waves in the dual geometries  means that they necessarily depend non-trivially on five of the six dimensions. The construction of these geometries is only possible because of the dramatic simplification afforded by the linear structure of the BPS equations and the decomposition of the solution into its ``linear pieces'' \cite{Bena:2011dd}.  Once these pieces are reassembled into the complete geometry, the metric appears to be remarkably complex, as it must be to encode all the physical data of the underlying CFT states.    

One of the remarkable features that has become evident in recent constructions of asymptotically-AdS superstrata  \cite{Bena:2016ypk,Bena:2017upb,Heidmann:2019xrd} is that most of the interesting physics of superstrata is encoded in a three-dimensional space-time, $\cK$.  Indeed,   the six-dimensional space-time of a  superstratum naturally decomposes into the $S^3$ surface around the branes, the radial coordinate, $r$, and the common directions, $(t,y)$ along the branes.\footnote{The remaining directions of the D5 are compactified on the $\IT^4$ that reduces IIB supergravity to six dimensions.}  The manifold, $\cK$, is the geometry described by the coordinates $(t, y, r)$  that are complementary to the $S^3$.  

Since we are working with the holographic dual of a $(1+1)$-dimensional CFT, the geometry is asymptotic to AdS$_3$ $\times S^3$, and the vacuum is  simply global AdS$_3$ $\times S^3$.   Superstrata involve turning on new  fluxes and adding metric deformations, thereby creating a warped, fibered product, $\cK \times S^3$.  The manifold, $\cK$, is a smooth, horizonless, three-dimensional space-time and the $S^3$ is usually deformed and fibered over $\cK$ by  non-trivial Kaluza-Klein Maxwell fields. 
 
The manifold, $\cK$, is best described as a ``smoothly-capped BTZ geometry.''  That is, like BTZ, it  is asymptotic to  AdS$_3$ at infinity and has a long AdS$_2$ $\times S^1$ throat, but unlike BTZ, this throat has a finite depth because it caps off smoothly without a horizon. These geometries thus look much like the horizon region of a black hole, except that there there is a finite redshift between the cap and any point in the asymptotic region.  It is these three-dimensional geometries that have provided the basis of many of the recent studies and comparison between microstate geometries and black holes \cite{Tyukov:2017uig,Raju:2018xue,Bena:2018mpb,Bena:2019azk,Heidmann:2019zws}.   

The analysis of such microstate geometries was greatly facilitated by the fact that, for some superstrata, the massless scalar wave equation in six dimensions is separable \cite{Bena:2017upb,Walker:2019ntz,Heidmann:2019zws}, reducing to a simple Laplacian on a ``round'' $S^3$ and a far more complicated wave equation on $\cK$.  For these geometries, the physics of massless scalar waves could indeed be entirely reduced to a problem on $\cK$.  It was also conjectured, based on indirect evidence, in \cite{Bena:2017upb} that some  superstrata should be part of a consistent truncation to a gauged supergravity in three dimensions. 

The purpose of this paper is to prove this conjecture by showing that the six-dimensional gauged supergravity that is the ``work-horse'' of superstrata construction, does indeed have a consistent truncation down to a three-dimensional gauged supergravity.  We will also give some explicit superstrata solutions that are entirely captured by this  truncation.

 Consistent truncations have a long history in supergravity and we will not review this here.  There are the relatively trivial consistent truncations that are based on reducing a higher-dimensional supergravity on a manifold that has isometries and restricting fields to singlets of those isometries.  This includes all the standard torus compactifications.  There are also highly non-trivial consistent truncations that involve sphere compactifications in which one keeps higher-dimensional fields that depend (at linear order) on particular sets of ``lowest harmonics'' on the sphere.  These fields therefore, typically, transform non-trivially under the rotation group of the sphere.  The isometries of the sphere also give rise to a non-abelian gauge symmetry in the lower dimension.   The end result  is a compactification that reduces a sector of the higher-dimensional supergravity to gauged supergravity in lower dimensions.  Here we will be concerned with $S^3$ compactifications of six-dimensional supergravity coupled to some tensor multiplets, and the corresponding three-dimensional gauged supergravity theory. We will also show the consistent truncation encodes some rich families of superstrata, some of which have been constructed elsewhere \cite{Bena:2017upb,Heidmann:2019xrd}.
  
We also construct new families of superstrata that depend on two freely-choosable holomorphic functions and that live entirely within our  consistent truncation. 
 
 An important point about consistent truncations is that they are not merely lower-dimensional effective  field theories.   If one solves the lower-dimensional equations of motion in a consistent truncation, the result is an {\it exact} solution of the higher-dimensional equations of motion.  This fact can be immensely useful in simplifying the equations of motion.  In particular, the sphere becomes an ``auxiliary'' space whose dynamics is entirely determined by the lower-dimensional theory and encoded in the details of the consistent truncation.  In this way, one can reduce a higher dimensional problem to a much more tractable lower-dimensional problem. 

Consistent truncations can prove to be a `Faustian Bargain.'  The price of the simplification is a huge restriction on the degrees of freedom: the higher dimensional theory has vastly more degrees of freedom than the lower-dimensional theory and these extra degrees of freedom may prove essential to capturing the correct physics.  The study of holographic $(2+1)$- and $(3+1)$-dimensional  field theories is littered with examples in which  consistent truncations have captured the essential physics, as well as examples in which the consistent truncation has lacked the necessary resolution to produce the correct physics.    We will discuss this further in Section \ref{sec:Conclusions}.

We have several reason for constructing the consistent truncations that are relevant to superstrata.  

First, motivated by the success of such a strategy for holographic field theories in $(2+1)$ and $(3+1)$ dimensions,  we wish to mine everything that  three-dimensional supergravities have to tell us about holographic field theories in $(1+1)$-dimensions, and the corresponding supergravity solutions in   in six dimensions.  Again, the lower dimensional BPS equations are much simpler than the higher-dimensional BPS equations, since solutions are functions of 2 rather than 5 variables, and thus may yield extremely interesting new holographic flows.  The three-dimensional formulation may also lead to a deeper understanding of the moduli space of superstrata and the microstates they represent. For example, we know, from perturbation theory \cite{Ceplak:2018pws,Tyukov:2018ypq}, that there are supersymmetric metric perturbations of superstrata.  As yet, we do not know how to ``integrate'' these perturbations up to finite moduli and thereby create new families of superstrata.  It is possible that the three-dimensional formulation will simplify a class of these moduli and show us how to do this more generally. 

Above all, is the possibility of getting a handle on non-supersymmetric, non-BPS superstrata. 

Given the intrinsic complexity of even the supersymmetric superstrata in six-dimensions, it seems an overwhelming task to address the non-linear equations that necessarily underlie the construction of non-BPS superstrata.   Indeed, such generic non-BPS superstrata are expected to depend non-trivially on all six dimensions. However,  the consistent truncation we present in this paper reduces this problem, for some limited families of superstrata, to a three-dimensional problem.   Solving the  equations of motion for the three-dimensional supergravity will still be a formidable task, and we intend to explore this in future work.  The importance of the results presented here is that they transform an impossible six-dimensional problem into a feasible three-dimensional problem.  

\subsection*{Paper overview}
In Section \ref{Sect:3Dsugr}, we describe the class of three-dimensional gauged supergravity theories that can encode superstrata; a summary of the supergravity theory, fields, and action is given in Section \ref{sec:3Dsummary}. The details of how this theory uplifts to six-dimensional supergravity may be found in Section \ref{Sect:uplift}.  Specifically,  we show how the consistent truncation works: how the three-dimensional fields are encoded in the six-dimensional supergravity and how the solutions of the three-dimensional equations yield a solution to the six-dimensional equations. In Section \ref{Sect:superstrata} we describe a new class of six-dimensional BPS superstrata (whose computational details may be found in Appendix \ref{sect:6Dsuperstrata}) that fit within the consistent truncation described in Section \ref{Sect:3Dsugr}.  We reduce these six-dimensional solutions to their three-dimensional data and use them to test the details of the consistent truncation.  The BPS superstrata that we have construct are  intrinsically new in that such a multi-function family, while in similar spirit to those in \cite{Heidmann:2019xrd}, have not been constructed before.

In Section \ref{sec:Conclusions} we make some final remarks and return to the discussion of the applications of our results.

\section{The  three-dimensional gauged supergravity}
\label{Sect:3Dsugr}

In this section, we will discuss a specific three-dimensional gauged supergravity theory which is relevant for the dimensional reduction of six-dimensional superstrata. The summary of our resulting three-dimensional theory is given in Section \ref{sec:3Dsummary}.

\subsection{Some supergravity background}
\label{sec:background}

If one reduces IIB supergravity on $\IT^4$ , one obtains the $\cN = (2,2)$ theory in six dimensions.  Reducing on a $K3$, instead, halves the supersymmetry to those that are holonomy invariant, and the result is an $\cN = (2,0)$ supergravity theory coupled to 21 anti-self-dual tensor multiplets.  

More generally, the ``parent theories'' of interest here are six-dimensional $\cN = (2,0)$ supergravity (with sixteen supersymmetries) coupled to $n$ tensor multiplets.  In such theories, the graviton multiplet contains one graviton, two complex, left-handed gravitinos (or four symplectic-Majorana Weyl gravitinos) and five self-dual, rank-two tensors gauge fields.  Each tensor multiplet contains one anti-self-dual, rank-two tensor gauge fields, two right-handed complex spinors (or four symplectic-Majorana Weyl spinors)  and five real scalars.  The $\cR$-symmetry is $SO(5) \cong USp(4)$, the tensor gauge fields transform in the fundamental of $SO(5,n)$ and the scalars are described in terms of a coset:
\begin{equation}
\frac{SO(5,n)}{SO(5) \times SO(n)} \,.
\label{coset1}
\end{equation}
In the fully non-linear theory, the scalar matrix plays an essential role in a twisted duality condition on the tensor gauge fields.  We will discuss a reduced version of this below. 

This six-dimensional supergravity can then be compactified on AdS$_3$ $\times S^3$ using a `Freund-Rubin' Ansatz in which one of the self-dual field strengths is set equal to the volume form of AdS$_3$ and of $S^3$.  This corresponds to the D1-D5 background in which the supergravity charges, $Q_1$ and $Q_5$, are set equal. If one wants unequal charges one must move some flux into an anti-self-dual tensor gauge field.  The simple, self-dual flux  breaks the $SO(5,n)$ symmetry to $SO(4,n)$.

There is now an extensive literature \cite{Cvetic:2000dm,Cvetic:2000zu, Nicolai:2001ac, Nicolai:2003bp, Nicolai:2003ux,Deger:2014ofa,Samtleben:2019zrh} on how this compactification leads to $\Neql8$ ($16$ supersymmetries) gauged supergravity in three dimensions. The gauge group is $SO(4) \cong SO(3)_+ \times SO(3)_-$ and comes from the isometries of $S^3$; the scalar coset becomes\footnote{Here we are going to consider the lowest KK towers in the compactification.  Remarkably, it seems  that one can consistently truncate in a manner that allows higher modes in the KK towers \cite{Nicolai:2003ux}.  As we will discuss in Section \ref{sec:Conclusions},  this might prove immensely useful in using three-dimensional supergravity to construct much more general classes of superstrata.}:
\begin{equation}
\frac{SO(8,3+n)}{SO(8) \times SO(3+n)} \,.
\label{coset2}
\end{equation}
The gauge group sits inside $SO(8,n+3)$ as the diagonal $SO(4)$ in the first and third  factors  of the decomposition \cite{Nicolai:2003ux}:
\begin{equation}
SO(4) \times SO(4) \times SO(4) \times SO(n-1)  \subset ~SO(8,3+n)  \,.
\label{groups}
\end{equation}
In particular, the precise relationship between   $\cN = (2,0)$ supergravity coupled to one anti-self-dual tensor multiplet in six dimensions and the  three-dimensional $\Neql8$, $SO(8,4)$ supergravity was recently laid out in \cite{Samtleben:2019zrh}.

The construction of superstrata  usually takes place in the less supersymmetric,  $\cN = (1,0)$ theories in six dimensions.   In such theories, the graviton multiplet contains one graviton, one complex, left-handed gravitino (or two symplectic-Majorana Weyl gravitinos) and one self-dual, rank-two tensors gauge field.   The tensor multiplet contains one anti-self-dual, rank-two tensor gauge fields, one right-handed complex spinor (or two symplectic-Majorana Weyl spinors) and one real scalar.   

The  simplest version of the theory used in superstratum construction can be characterized \cite{Bena:2015bea} as taking the bosonic fields to be those  obtained by making a $\IT^4$ compactification of IIB supergravity and then further restricting to only the fields that transform trivially under the $SO(4)$ global rotations on the tangent space of the $\IT^4$.  This results in $\cN = (1,0)$ supergravity coupled to two anti-self dual tensor multiplets.  To be precise, the ten-dimensional RR field, $C^{(2)}$, descends to the self-dual tensor in the gravity multiplet and one of the two anti-self-dual tensors.  These two components are independent and account for the separate D1 and D5 pieces.  The other anti-self-dual tensor descends from the ten-dimensional Kalb-Ramond field, $B^{(2)}$, and anti-self-duality is required by supersymmetry.  Roughly, the only way that this can be compatible with $\cN = (1,0)$ supersymmetry in a D1-D5 system is if the F1 and NS5 fields are locked together via anti-self-duality.

  To go from the $(2,0)$ supergravity theory to the $(1,0)$ theory,  one must remove two $(1,0)$ gravitino multiplets, which include four of the five self-dual tensor gauge fields. One also removes a corresponding number of hypermultiplets.  This reduces the  $SO(5,n)$ symmetry,  and the scalar coset,  of the six-dimensional theory  to $SO(1,n)$. The compactification to three dimensions on AdS$_3$ $\times S^3$ then results in the scalar coset
\begin{equation}
\frac{SO(4,3+n)}{SO(4) \times SO(3+n)} \,.
\label{coset3}
\end{equation}
The  relationship between  $\cN = (1,0)$ supergravity coupled to one anti-self-dual tensor multiplet in six dimensions and the  three-dimensional $\Neql4$, $SO(4,4)$ supergravity (with $n=1$) has been laid out in \cite{Cvetic:2000dm,Cvetic:2000zu,Cvetic:2000ah, Deger:2014ofa}.

As we noted above, superstrata require $\cN = (1,0)$ supergravity coupled to at least two tensor multiplets. 
The extra tensor multiplet plays an essential role in the construction of {\it smooth} solutions. Indeed, string amplitude computations \cite{Giusto:2011fy,Giusto:2012jx} and the holographic dictionary showed that it is inconsistent to freeze out this degree of freedom.  It was this realization that led to the first successful construction of non-trivial superstrata in \cite{Bena:2015bea}.  

We also note that the three-dimensional supergravity corresponding to $\cN = (1,0)$ supergravity coupled to {\it one} tensor multiplet was used in \cite{Deger:2019jtl} to construct some  black-ring and black-string solutions.  The consistent truncation proved to be a useful tool, but the lack of the extra tensor multiplet meant that the solutions were singular and that superstrata were inaccessible from within such a truncation.

We are therefore going to examine the  three-dimensional, $(0,2)$ supergravity (with eight supersymmetries) for which the scalar coset is 
\begin{equation}
\frac{SO(4,5)}{SO(4) \times SO(5)} \,.
\label{coset4}
\end{equation}
As will become evident, we will find the results in  \cite{Samtleben:2019zrh} for $\cN = (2,0)$ supergravity immensely useful in extending the results of  \cite{Deger:2014ofa} to obtain the three-dimensional supergravity corresponding to $\cN = (1,0)$ supergravity coupled to two tensor multiplets.


The relevant supergravity in three dimensions is fully defined by its amount of supersymmetry, the scalar coset, the gauge symmetry and the gauge couplings as defined by an embedding tensor.   The  number of bosonic degrees of freedom in the theory is equal to the dimension of  the underlying coset.  However, these degrees of freedom can be encoded in various ways in the action.  In three dimensions, Yang-Mills gauge fields can be dualized into scalars and vice versa and this is how the Yang-Mills fields can be generated.   In addition, one of the essential features of the three-dimensional theories is the appearance of massive Chern-Simons vector fields.  These fields can be viewed as gauging non-semi-simple groups and can ultimately be integrated out. Thus, as explained in \cite{Nicolai:2003bp}, the number of bosonic degrees of freedom, $d$, is given by: 
\begin{equation}
d~=~ {\rm dim}({\rm Coset})~=~ \# ({\rm Scalars}) ~+~  \# ({\rm YM \ vectors}) ~+~  \# ({\rm massive \ CS \ vectors})\,.
\label{bosedofs}
\end{equation}
For the  three-dimensional $(0,2)$ supergravity theories described above, with coset (\ref{coset3}), the gauge group is actually a semi-direct product $SO(4) \ltimes \IT^6 \subset SO(4,3+n)$, where $SO(4)$ is the standard  Yang-Mills gauge group coming from $S^3$ and $\IT^6$ is a translation that transforms in the adjoint of $SO(4)$.  Thus the $4\times (3+n)$ degrees of freedom become $6$ YM vectors, $6$ CS vectors and $4n$ scalars.

\subsection{The scalar degrees of freedom}
\label{sec:scalars}

To describe this theory we simply follow the discussion in \cite{Deger:2014ofa} but with an extra tensor multiplet.  The $SO(4,5)$ group has an invariant metric: 
\begin{equation}
\eta ~\equiv~
\left( \begin{matrix} 
0_{4 \times 4} & \oneone_{4 \times 4} & 0 \\
\oneone_{4 \times 4} & 0_{4 \times 4} & 0 \\
0 &0& \varepsilon
\end{matrix} \right) \,,
\label{invmat}
\end{equation}
where $\varepsilon = \pm 1$.  Note that we are using the anti-diagonal form of $\eta$ because it is far more convenient in describing the degrees of freedom and in expressing the gauging. For $\varepsilon =-1$ we have $G= SO(5,4)$ (in the conventions of \cite{Samtleben:2019zrh}) and for  $\varepsilon =1$ we have $G= SO(4,5)$.  While this might seem a trivial notational distinction, it is related to the self-duality or anti-self-duality of the additional tensor multiplet.  The theory of interest to us to descrive superstrata has $\varepsilon =1$, whereas (a truncated version of) the theory in \cite{Samtleben:2019zrh} corresponds to $\varepsilon =-1$.

The generators of $G$ may be written as:
\begin{equation}
\left( \begin{matrix} 
A  & B & \chi_A \\
C & -A^T & \lambda^A  \\
-\varepsilon  \lambda^A &-\varepsilon  \chi_A & 0
\end{matrix} \right) \,,  
\end{equation}
where $A,B,C,D$ are $4 \times 4$ matrices with $B^T = -B$ and $C^T = -C$. The matrix $A$ generates a $GL(4,\IR)$ whose compact generators define the $SO(4)$ YM gauge group and whose non-compact generators are obtained by taking $A^T=A$.   The remaining $10$ non-compact generators can be taken to be $B$ and $\chi_A$.  That is, we will choose to parametrize the  coset  by setting $A^T=A$, $C=0$ and $\lambda^A =0$.  The matrix $B$ describes the translation generators of $\IT^6$  transforming in the adjoint of $SO(4)$.   In this formulation, the $20$ bosonic degrees of freedom are defined by $A^T=A$, $B^T = -B$ and $\chi_A$.  

The simplest way to fix the $\IT^6$ gauge invariance is  to set $B=0$, which we will now do. We could also fix the $SO(4)$ gauge invariance by reducing $A$ to a diagonal matrix.  The $20$ degrees of freedom would then be the $4$ eigenvalues, the $4$ $\chi_A$'s and $6+6$ gauge fields.  However, we will only go half-way: fixing the $\IT^6$ gauge and moving these degrees of freedom into the CS vectors. We will preserve the $SO(4)$ gauge invariance.

Thus our scalar matrix will be defined by:
\begin{equation}
\begin{aligned}
{{\cal V}_{\bar M}}^{\bar K} & ~=~
 \exp \left( \begin{matrix} 
0 & 0& \chi_A \\
0& 0 & 0 \\
0 &-\varepsilon  \chi_A & 0
\end{matrix} \right) \, \left( \begin{matrix} 
{P_A}^B & 0& 0 \\
0& {(P^{-1})_B}^A & 1 \\
0 &0& 0
\end{matrix} \right)    \\
& ~=~
\left( \begin{matrix} 
{P_A}^B & - \frac{1}{2}\varepsilon\, \chi_A \,  \big({(P^{-1})_B}^C \chi_C\big) & \chi_A  \\
0& {(P^{-1})_B}^A & 0 \\
0 &-\varepsilon{(P^{-1})_B}^C \chi_C & 1
\end{matrix} \right)  \,,
\end{aligned}
\label{scalmat}
\end{equation}
where $P =P^T$ is a symmetric $GL(4,\IR)$ matrix. 

Our index conventions will be as follows.  A vector of $G$ will be denoted by 
\begin{equation}
\cX_{\bar M}  ~\equiv~ ( \cX_A, \cX^A, \cX_0) \,,   \qquad X^{\bar M}  ~\equiv~ ( \cX^A, \cX _A, \varepsilon \cX_0) \,,
\end{equation}
where the indices are raised and lowered using (\ref{invmat}).  The components, $X_A$ and $X^A$, transform, respectively, in the $4$ and $\overline 4$ of $GL(4,\IR)$.  That is, they transform through multiplication by $P$ or $P^{-1}$, respectively.

Following \cite{Samtleben:2019zrh}, we define:
\begin{equation}
\begin{aligned}
 M_{\bar A \bar B}  ~=~  & ({\cal V} {\cal V}^T)_{\bar A \bar B}    ~=~ {{\cal V}_{\bar A}}^{\bar C}  {{\cal V}_{\bar B}}^{\bar C}   \,, \qquad
  M^{\bar A \bar B}  ~=~   \big(({\cal V}^T)^{-1} \, ({\cal V}^{-1} )\big)^{\bar A \bar B}    ~=~\big({ {\cal V}^{-1}\big)_{\bar C}}^{\bar A}   \big({ {\cal V}^{-1}\big)_{\bar C}}^{\bar B} \,,\\
 m_{A  B}  ~=~   & (P \, P^T)_{AB}    ~=~{P_A}^C \, {P_B}^C \,, \qquad    m^{A  B}  ~=~    \big( (P^T)^{-1} P^{-1}\big)^{AB}    ~=~{(P^{-1})_C}^A\, {(P^{-1})_C}^B \,.
\end{aligned}
\label{scalarmats}
\end{equation}
%

\subsection{The gauge couplings}
\label{sec:gaugecouplings}

The embedding of the gauge group $SO(4) \ltimes \IT^6$ in $G$ is defined through the embedding tensor $\Theta$.  Specifically, if $T^{\bar M \bar N}=-T^{\bar N \bar M}$ are the generators of $G$, then the covariant derivative is defined by:
\begin{equation}
\widehat\cD_\mu \, \cX_{\bar P}  ~\equiv~ \partial_\mu \, \cX_{\bar P}  ~+~ {A_\mu}^{\bar K \bar L} \, \Theta_{\bar K \bar L, \bar M \bar N } \, (T^{\bar M \bar N})_{\bar P}{}^{\bar Q} (\cX_{\bar Q} )   \,, 
\label{covderiv1}
\end{equation}
where ${A_\mu}^{\bar K \bar L}$ are the gauge connections and $(T^{\bar M \bar N})_{\bar P}{}^{\bar Q} (\cX_{\bar Q} )$ represents the action of $T^{\bar M \bar N}$ on vectors.  In the standard normalization, one has:
\begin{equation}
 (T^{\bar M \bar N})_{\bar P}{}^{\bar Q} (\cX_{\bar Q} ) ~=~ \delta^{\bar N}_{\bar P} \cX^{\bar M} ~-~ \delta^{\bar M}_{\bar P} \cX^{\bar N}   \,, 
\label{Tform}
\end{equation}
and these matrices have the commutation relations\footnote{Our conventions, and the signs of the structure constants, differ from those of \cite{Samtleben:2019zrh}.  We discuss our choices and consistent formulation of the gauge action in detail in   Appendix \ref{sec:app:gauge}.}: 
\begin{equation}
\begin{aligned}
\big[ \, T^{\bar K \bar L}\,, T^{\bar M \bar N} \,\big] ~=~ &  f^{\bar K\bar L,\bar M\bar N}{}_{\bar P\bar Q}\, T^{\bar P \bar Q} \\
 ~=~ &   \eta^{\bar K \bar M} T^{\bar L \bar N}~+~ \eta^{\bar K \bar N} T^{\bar M \bar L } ~-~\eta^{\bar L \bar M} T^{\bar K \bar N}~-~ \eta^{\bar L \bar N} T^{\bar M \bar K }\,.
\end{aligned}
\label{comms}
\end{equation}
This defines the structure constants: 
\begin{equation}
\begin{aligned}
f^{\bar K\bar L,\bar M\bar N}{}_{\bar P\bar Q}  ~=~ &
  \coeff{1}{2}\, \eta^{\bar K \bar M} \, \big( \delta^{\bar L}_{\bar P} \,\delta^{\bar N}_{\bar Q}   - \delta^{\bar L}_{\bar Q} \,\delta^{\bar N}_{\bar P}\big)+~  \coeff{1}{2}\,\eta^{\bar K \bar N} \, \big( \delta^{\bar M}_{\bar P} \,\delta^{\bar L}_{\bar Q}   - \delta^{\bar M}_{\bar Q} \,\delta^{\bar L}_{\bar P}\big) \\
  & ~-~ \coeff{1}{2}\,\eta^{\bar L \bar M} \, \big( \delta^{\bar K}_{\bar P} \,\delta^{\bar N}_{\bar Q}   - \delta^{\bar K}_{\bar Q} \,\delta^{\bar N}_{\bar P}\big) ~-~  \coeff{1}{2}\, \eta^{\bar L \bar N} \, \big( \delta^{\bar M}_{\bar P} \,\delta^{\bar K}_{\bar Q}   - \delta^{\bar M}_{\bar Q} \,\delta^{\bar K}_{\bar P}\big)
\end{aligned}
\label{structureconsts}
\end{equation}
in which indices are summed without any weight factors\footnote{This ``double counts'' the generators because $T^{\bar P \bar Q} =-T^{\bar Q \bar P}$.  This is, however, a completely standard convention that we use everywhere in this paper.}. 

The generic form of the embedding tensor is:
\begin{equation}
\Theta_{\bar K \bar L, \bar M \bar N  } ~=~ \theta_{\bar K \bar L  \bar M \bar N } ~+~ \coeff{1}{2}\, \big( \eta_{{\bar M}[{\bar K}}\, \theta_{{\bar L}]{\bar N}} -\eta_{{\bar N}[{\bar K}}\, \theta_{{\bar L}]{\bar M}}  \big) ~+~ \theta \, \eta_{{\bar M}[{\bar K}}\, \eta_{{\bar L}]{\bar N}}\,,
\end{equation}
where $ \theta_{[\bar K \bar L  \bar M \bar N ]} = \theta_{\bar K \bar L  \bar M \bar N } $ and $ \theta_{\bar K \bar L} = \theta_{\bar L \bar K } $.  However, for the gauged  $SO(4) \times \IT^6$ theory of interest here, the only non-vanishing pieces are \cite{Nicolai:2003ux,Deger:2014ofa,Samtleben:2019zrh}:   
\begin{equation}
 \theta_{A B CD }  ~=~ -2\, \alpha \,   \epsilon_{A B CD }  \,, \qquad  {\theta_{A B C }}^D  ~=~ \gamma_0 \,   \epsilon_{A B C E} \, \delta^{DE}   \,,
\label{EmbTens}
\end{equation}
for some coupling constants $\alpha$ and $\gamma_0$.  It is in this expression that the $GL(4,\IR)$ formulation arising from the choice (\ref{invmat}) leads to significant simplification.

In particular, this embedding tensor reduces ${A_\mu}^{\bar K \bar L}$ to the twelve independent gauge fields for $SO(4) \ltimes \IT^6$:
\begin{equation}
 {A_\mu}^{AB} ~=~ - {A_\mu}^{BA}   \,, \qquad  A_{\mu}{}^{B}{}_A ~=~  -{A_{\mu \,B}}^A \,.
\end{equation}
It is convenient to define:
\begin{equation}
{\widetilde A_\mu}{}^{AB}  ~\equiv~ \coeff{1}{2} \,\epsilon_{ABCD}\,{A_\mu}^{CD} \,, \qquad \ {\widehat A_\mu}{}^{AB}  ~\equiv~ \coeff{1}{2} \,\epsilon_{ABCD}\,{{A_\mu}{}^C}{}_D   \,, 
\label{dualGFs}
\end{equation}
and introduce:
\begin{equation}
{B_\mu}{}^{AB}  ~\equiv~ 8\, \big(\alpha\, {\widetilde A_\mu}{}^{AB}    - \gamma_0 \, {\widehat A_\mu}{}^{AB} \big)  \,.
\label{Bvecden}
\end{equation}

One then finds that (\ref{covderiv1}) can be written in terms of $GL(4,\IR)$ components as:
\begin{equation}
\begin{aligned}
\widehat\cD_\mu \, \cX_{A}  &~=~ \partial_\mu \, \cX_{A}  ~+~ {B_\mu}{}^{AB}  \,  \cX^{B}  ~-~  4\, \gamma_0\,\widetilde A_\mu{}^{AB} \, \cX_{B} \,, \\  
\widehat\cD_\mu \, \cX^{A}  &~=~ \partial_\mu \, \cX^{A}    ~-~  4\, \gamma_0\,\widetilde A_\mu{}^{AB} \, \cX^{B} \,, \qquad \widehat\cD_\mu \, \cX_0 ~=~ \partial_\mu \, \cX_0  \,.
\end{aligned}
\label{covderiv2}
\end{equation}
Note that, in terms of the matrices of $G$, the connection ${B_\mu}{}^{AB}$ has the form:
\begin{equation}
\cB_\mu ~\equiv~
\left( \begin{matrix} 
0 & {B_\mu}{}^{AB} & 0\\
0 & 0 & 0 \\
0 &0 & 0
\end{matrix} \right) \,.
\end{equation}
These are therefore  precisely the gauge fields of $\IT^6$.  The vector fields $A_\mu{}^{AB}$ are those of $SO(4)$ but they act with their duals, and with a gauge coupling of $-4 \gamma_0$. 

To make this more explicit, define the  $SO(3)_+ \times SO(3)_-$ parts of the gauge connection, and its dual:
\begin{equation}
 A_\mu{}^{AB}  ~=~  A^+_\mu{}^{AB} ~+~  A^-_\mu{}^{AB} \,, \qquad  \widetilde  A_\mu{}^{AB}  ~=~ A^+_\mu{}^{AB} ~-~  A^-_\mu{}^{AB}    \,,
\label{decomp}
\end{equation}
and define the gauge couplings 
\begin{equation}
g_+  ~=~ -g_- ~=~ -4 \gamma_0 \,.
\label{gpmdefn}
\end{equation}
Then one has
\begin{equation}
\begin{aligned}
\widehat\cD_\mu \, \cX_{A}  &~=~ \partial_\mu \, \cX_{A}  ~+~ {B_\mu}{}^{AB}  \,  \cX^{B} ~+~   g_+ A^+_\mu{}^{AB}  \, \cX_{B} ~+~   g_- A^-_\mu{}^{AB}  \, \cX_{B}  \,, \\  
\widehat\cD_\mu \, \cX^{A}  &~=~ \partial_\mu \, \cX^{A}    ~+~   g_+ A^+_\mu{}^{AB}  \, \cX^{B} ~+~  g_- A^-_\mu{}^{AB}  \, \cX^{B}\,, \qquad \widehat \cD_\mu \, \cX_0 ~=~ \partial_\mu \, \cX_0  \,.
\end{aligned}
\label{covderiv3}
\end{equation}

Finally, it is convenient to define the reduced, purely-$SO(4)$, covariant derivatives:
\begin{equation}
\begin{aligned}
\cD_\mu \, \cX_{A}  &~=~ \partial_\mu \, \cX_{A}   ~-~  4\, \gamma_0\,\widetilde A_\mu{}^{AB} \, \cX_{B} \,, \\  
 \cD_\mu \, \cX^{A}  &~=~ \partial_\mu \, \cX^{A}    ~-~  4\, \gamma_0\,\widetilde A_\mu{}^{AB} \, \cX^{B} \,, \qquad  \cD_\mu \, \cX_0 ~=~ \partial_\mu \, \cX_0  \,.
\end{aligned}
\label{covderiv4}
\end{equation}
%

\subsection{The scalar action}
\label{sec:scalaraction}

From \cite{Samtleben:2019zrh}, the scalar action is
\begin{equation}
{\cal L}_{\rm scalar} ~=~  \frac{1}{32}\, \big( {\widehat\cD_\mu} M_{\bar K \bar L} \big) \, \big( {\widehat\cD^\mu} M^{\bar K \bar L}\big)  ~-~ V \,,
\end{equation}
where the potential, $V$, is given by:
\begin{equation}
\begin{aligned}
V  ~=~&  \frac{1}{48} \,  \theta_{\bar K\bar L\bar M\bar N}\, \theta_{\bar P\bar Q\bar R\bar S} \, \Big(M^{\bar K\bar P} M^{\bar L\bar Q} M^{\bar M\bar R} M^{\bar N\bar S}
~-~ 6\, M^{\bar K\bar P} M^{\bar L\bar Q} \eta^{\bar M\bar R} \eta^{\bar N\bar S}  \\
& \qquad\qquad\qquad \qquad\qquad ~+~ 8 \, M^{\bar K\bar P} \eta^{\bar L\bar Q} \eta^{\bar M\bar R} \eta^{\bar N\bar S}
~-~ 3 \, \eta^{\bar K\bar P} \eta^{\bar L\bar Q} \eta^{\bar M\bar R} \eta^{\bar N\bar S} \Big)  \\
&~+~  \frac{1}{32} \, \theta_{\bar K\bar L} \theta_{\bar P\bar Q} \Big(2 M^{\bar K\bar P} M^{\bar L\bar Q} - 2 \eta^{\bar K\bar P} \eta^{\bar L\bar Q}
~-~ M^{\bar K\bar L} M^{\bar P\bar Q} \Big)   ~+~ \theta \, \theta_{\bar K\bar L} M^{\bar K\bar L} ~-~8\, \theta^2 
\end{aligned}
\end{equation}
Using the expressions above, we find the following result:
\begin{equation}
\begin{aligned}
{\cal L}_{\rm scalar} ~=~ & -\coeff{1}{16}\,  {\rm Tr}\big[ \big( \cD_\mu  m \big) m^{-1}\, \big( \cD^\mu m \big) m^{-1}\,  \big]  -\coeff{1}{8}\, m^{AB} \,(\cD_\mu \chi_A)\,(\cD^\mu \chi_B)\\
 &-\coeff{1}{16}\, m^{AC} \,m^{BD}\,\Big({B^\mu}{}^{AB} - \coeff{1}{2}\, \varepsilon \,Y^\mu _{  AB} \Big) \,\Big({B_\mu}{}^{CD} - \coeff{1}{2} \,\varepsilon\, Y_{\mu \, CD} \Big) ~-~ V \,.
\end{aligned}
\end{equation}
where $\cD_\mu$ is the $SO(4)$ covariant derivative defined in (\ref{covderiv4}), and where 
\begin{equation}
Y_{\mu \, AB}  ~\equiv~  \chi_B \,\cD_\mu \chi_A ~-~  \chi_A \,\cD_\mu\chi_B \,,
\end{equation}
and
\begin{equation} \label{eq:3Dpotential}
V  ~=~   \det\big(m^{AB}\big)\, \Big[\, 2\,\big(\alpha ~+~ \coeff{1}{4} \,\varepsilon  \, \gamma_0 (\chi_A \chi_A)\big)^2 ~+~ \gamma_0^2 \, \big( m_{AB}\,\big( m_{AB} ~+~ \coeff{1}{2} \, \chi_A \chi_B \big)~-~ \coeff{1}{2}m_{AA}\, m_{BB}   \big) \Big]   \,.
\end{equation}
%

\subsection{The Chern-Simons action}
\label{sec:CSaction}

The general Chern-Simons term is:
\begin{equation}
\cL_{CS} ~=~  \frac{1}{4}\, \varepsilon^{\mu \nu \rho} \,  A_\mu{}^{\bar K\bar L} \,\Theta_{\bar K\bar L,\bar M\bar N}\,
\Big( \partial_\nu A_\rho {}^{\bar M\bar N}  ~+~  \frac{1}{3}\, f^{\bar M\bar N,\bar P\bar Q}{}_{\bar R\bar S}\,\Theta_{\bar P\bar Q,\bar U\bar V}\,A_\nu{}^{\bar U\bar V} A_\rho{}^{\bar R\bar S}\Big)\,.
\label{CSgeneric}
\end{equation}
We have reversed the sign of the last term relative to \cite{Samtleben:2019zrh} so as to have the canonical Chern-Simons action.  (We discuss our choices and consistent formulation of the gauge action in detail in   Appendix \ref{sec:app:gauge}.)

Using the embedding tensor (\ref{EmbTens}) and the structure constants (\ref{structureconsts}), we find
\begin{equation}
\cL_{CS}  ~=~  -\varepsilon^{\mu \nu \rho} \, \Big[  \alpha\,\big(A_\mu{}^{AB}\, \partial_\nu  \widetilde A_\rho{}^{BA}  ~-~\coeff{8}{3}\,  \gamma_0 \, A_\mu{}^{AB} \,  A_\nu{}^{BC}\, A_\rho{}^{CA} \,\big) ~-~  \coeff{1}{8}\,  {B_\mu}{}^{BA}  \, F_{\nu \rho}^{AB} \Big] \,,
\label{CSterm}
\end{equation}
where:
\begin{equation}
F_{\nu \rho}^{AB}   ~\equiv~  2\,\big( \partial_{[\nu}   A_{\rho]}{}^{AB}   ~-~4 \,  \gamma_0\,   A_{[\nu} {}^{C[A} \,\widetilde  A_{\rho]} {}^{B]C}\big) \,.
\label{Maxdefn}
\end{equation}
When written in terms of  $SO(3)_+$ and $SO(3)_-$, we note that this action takes the more familiar Chern-Simons form:
\begin{equation}
\begin{aligned}
\cL_{CS} ~=~  -  \alpha\,\varepsilon^{\mu \nu \rho} \, \Big[  &\big(A^+_\mu{}^{AB}\, \partial^{}_\nu   A^+_\rho{}^{BA}  ~+~\coeff{2}{3}\,  g_+ \, A^+_\mu{}^{AB} \, A^+_\nu{}^{BC}\, A^+_\rho{}^{CA} \,\big) \\
 &-  \big(A^-_\mu{}^{AB}\, \partial^{}_\nu   A^-_\rho{}^{BA} ~+~\coeff{2}{3}\,  g_-  \, A^-_\mu{}^{AB} \, \widetilde A^-_\nu{}^{BC}\,\widetilde A^-_\rho{}^{CA} \,\big)  \Big]  ~+~  \coeff{1}{8}\,  \varepsilon^{\mu \nu \rho} \,{B_\mu}{}^{BA}  \, F_{\nu \rho}^{AB} \,,
 \end{aligned}
\label{CStermpieces}
\end{equation}
with:
\begin{equation}
F_{\nu \rho}^{AB}   ~=~  F^+{}_{\nu \rho}^{AB}  ~+~ F^-{}_{\nu \rho}^{AB}   \,, \qquad F^\pm{}_{\nu \rho}^{AB} ~\equiv~ 2\,\big( \partial^{}_{[\nu}   A_{\rho]}^\pm{}^{AB}   ~+~g_\pm\,   A^\pm_{[\nu} {}^{C [A } \,   A^\pm_{\rho]} {}^{B]C}\big) \,.
\label{SO3Fs}
\end{equation}
%

\subsection{Integrating out the Chern-Simons gauge fields}
\label{sec:CSintout}

The Chern-Simons gauge fields, ${B_\mu}{}^{AB}$, appear only quadratically in the action, and without derivatives.  It is therefore trivial to integrate them out by completing the square. 
The complete bosonic action may be written
\begin{equation}
\begin{aligned}
{\cal L}  ~=~ &\coeff{1}{4}\,R ~-~\coeff{1}{16}\,  {\rm Tr}\big[ \big( \cD_\mu  m \big) m^{-1}\, \big( \cD^\mu m \big) m^{-1}\,  \big]  -\coeff{1}{8}\, m^{AB} \,(\cD_\mu \chi_A)\,(\cD^\mu \chi_B) ~-~V \\
&- \coeff{1}{8} m_{AC} \,m_{BD}\, F_{\mu \nu}^{AB}\, F^{\mu \nu}{}^{CD}   ~-~ \alpha\,\varepsilon^{\mu \nu \rho} \, \big(A_\mu{}^{AB}\, \partial_\nu  \widetilde A_\rho{}^{BA}  - \coeff{8}{3}\,  \gamma_0 \, A_\mu{}^{AB} \,  A_\nu{}^{BC}\, A_\rho{}^{CA} \,\big)   \\ & ~-~\coeff{1}{16}\, \varepsilon \,  \varepsilon^{\mu \nu \rho} \, Y_\mu{}_{AB} \, F_{\nu \rho}^{AB} ~+~ {\cal L}_{\rm B} \,,
\end{aligned}
\label{action1}
\end{equation}
where
\begin{equation}
\begin{aligned}
{\cal L}_{\rm B} ~\equiv~ & -\coeff{1}{16}\, g_{\mu \nu} \, m^{AC} \,m^{BD}\,\Big({B^\mu}{}^{AB} + \varepsilon^{\mu  \rho_1 \rho_2} \, m_{AE_1} m_{BE_2}\,F_{\rho_1 \rho_2}^{E_1E_2} - \coeff{1}{2}\, \varepsilon \,Y^\mu{}_{ \, AB} \Big) \\
 &\qquad\qquad\qquad\qquad\qquad \times  \Big({B^\nu}{}^{CD} + \varepsilon^{\nu \rho_3 \rho_4}\, m_{CE_3} \,m_{DE_4}\,F_{\rho_3 \rho_4}^{  E_3 E_4}  - \coeff{1}{2} \,\varepsilon\, Y^\nu{}_{ CD} \Big) \,.
\end{aligned}
\label{Baction}
\end{equation}
Thus the equations of motion for ${B_\mu}{}^{AB}$ are trivial, and yield
\begin{equation}
{B^\mu}{}^{AB} ~=~ \coeff{1}{2}\, \varepsilon \,Y^\mu{}_{ \, AB}  ~-~  \varepsilon^{\mu  \nu \rho} \, m_{AC} m_{BD}\,F_{\nu \rho}^{CD}   \,,
 \label{Beqn}
\end{equation}
and these gauge fields drop out of the action entirely.

\subsection{The three-dimensional supergravity: summary and comments}
\label{sec:3Dsummary}

The three-dimensional supergravity theory we consider is a $(0,2)$ gauged supergravity with eight supersymmetries.  It has an $SO(4)$ gauge symmetry, with $6$ gauge fields, $A_\mu^{AB} = A_\mu^{[AB]}$, where the indices $A, B, \ldots  =1,2,3,4$ transform in the vector of $SO(4)$. In addition to the graviton, and the gauge fields, $A_\mu^{AB}$, there are $14$ scalar fields in the bosonic sector.  Four scalars are encoded in an $SO(4)$ vector, $\chi_A$, and the other ten are encoded as a general, symmetric $GL(4, \IR)$ matrix, $m_{AB}=m_{(AB)}$, with inverse, $m^{AB}$.

The bosonic action is  
\begin{equation}
\begin{aligned}
{\cal L}  ~=~ &\coeff{1}{4}\,R ~-~\coeff{1}{16}\,  {\rm Tr}\big[ \big( \cD_\mu  m \big) m^{-1}\, \big( \cD^\mu m \big) m^{-1}\,  \big]  -\coeff{1}{8}\, m^{AB} \,(\cD_\mu \chi_A)\,(\cD^\mu \chi_B) ~-~V \\
&- \coeff{1}{8} m_{AC} \,m_{BD}\, F_{\mu \nu}^{AB}\, F^{\mu \nu}{}^{CD}   ~-~ \alpha\,\varepsilon^{\mu \nu \rho} \, \big(A_\mu{}^{AB}\, \partial_\nu  \widetilde A_\rho{}^{BA}  - \coeff{8}{3}\,  \gamma_0 \, A_\mu{}^{AB} \,  A_\nu{}^{BC}\, A_\rho{}^{CA} \,\big)   \\ & ~-~\coeff{1}{16}\, \varepsilon \,  \varepsilon^{\mu \nu \rho} \, Y_\mu{}_{AB} \, F_{\nu \rho}^{AB} \,,
\end{aligned}
\label{action-final}
\end{equation}
where $m$ denotes $m_{AB}$, and 
\begin{equation}
Y_{\mu \, AB}  ~\equiv~  \chi_B \,\cD_\mu \chi_A ~-~  \chi_A \,\cD_\mu\chi_B \,.
\label{Ydefn}
\end{equation}
The covariant derivative is defined on upper and lower $SO(4)$ indices as
\begin{equation}
\cD_\mu \, \cX_{A}  ~=~ \partial_\mu \, \cX_{A} ~-~  4\, \gamma_0\,\widetilde A_\mu{}^{AB} \, \cX_{B} \,, \qquad 
 \cD_\mu \, \cX^{A}  ~=~ \partial_\mu \, \cX^{A}   ~-~   4\, \gamma_0\,\widetilde A_\mu{}^{AB} \, \cX^{B} \,,   
\label{covderiv-final}
\end{equation}
where 
\begin{equation}
{\widetilde A_\mu}{}^{AB}  ~\equiv~ \coeff{1}{2} \,\epsilon_{ABCD}\,{A_\mu}^{CD} \,.
\label{dualGFsfinal}
\end{equation}
The field strengths are given by
\begin{equation}
F_{\nu \rho}^{AB}   ~\equiv~  2\,\big( \partial_{[\nu}   A_{\rho]}{}^{AB}   ~-~4 \,  \gamma_0\,   A_{[\nu} {}^{C[A} \,\widetilde  A_{\rho]} {}^{B]C}\big) \,.
\label{Maxdefn-final}
\end{equation}
The scalar potential is:
\begin{equation}
V  ~=~   \det\big(m^{AB}\big)\, \Big[\, 2\,\big(\alpha ~+~ \coeff{1}{4} \,\varepsilon  \, \gamma_0 (\chi_A \chi_A)\big)^2 ~+~ \gamma_0^2 \, \big( m_{AB}\,\big( m_{AB} ~+~ \coeff{1}{2} \, \chi_A \chi_B \big)~-~ \coeff{1}{2}m_{AA}\, m_{BB}   \big) \Big]   \,.
 \label{eq:3Dpotential-final}
\end{equation}
The equations of motion following from (\ref{action-final}) are given explicitly in Appendix \ref{App:3Deom}.

The bosonic theory has three parameters:  the gauge coupling, $\gamma_0$,  a scale parameter, $\alpha$, and a ``signature,''  $\varepsilon = \pm1$.  In Appendix \ref{app:rescalings}, we discuss two scale invariances of the action that can be used to set $|\alpha| =1$ and $\gamma_0=1$.  We will, however, retain these parameters.

The parameter $\varepsilon$, and the sign of $\alpha$ are extremely important to the supersymmetry.  We will discuss this further in the six-dimensional context in Section \ref{sec:Establishing}.  Here we will simply note that sending  $\varepsilon \to - \varepsilon, \alpha \to - \alpha$  leaves the potential invariant.   In the rest of the action, the parameters $\varepsilon$ and $\alpha$ only appear as coefficients of the parity odd terms involving  $\varepsilon^{\mu \nu \rho}$.  Thus $\varepsilon \to - \varepsilon, \alpha \to - \alpha$, combined with an orientation reversal in the three-dimensional space-time is a symmetry of the action.  

\section{From three to six dimensions}
\label{Sect:uplift}

Here, we will describe how the three-dimensional theory of Section \ref{Sect:3Dsugr} can be obtained from a dimensional reduction from six dimensions. First, in Section \ref{sec:Establishing}, we discuss how we obtained our consistent truncation formulae. In Section \ref{sec:6Dtheorysub}, we describe the six-dimensional theory at hand, which is the theory relevant for the description of superstrata. In Section \ref{sec:full6Duplift}, we give the full non-linear reduction ansatz from this six-dimensional theory to the three-dimensional theory of Section \ref{Sect:3Dsugr}. Finally, in Section \ref{sec:U1sqtrunc}, we describe a useful $U(1)^2$ subsector of this truncation and its three-dimensional counterpart.

\subsection{Establishing the consistent truncation}
\label{sec:Establishing}

The common core of superstrata and the consistent truncations of \cite{Cvetic:2000dm,Cvetic:2000zu,Cvetic:2000ah, Deger:2014ofa, Samtleben:2019zrh}  is the basic  $\mathcal{N}=(1,0)$ supergravity coupled to single anti-self-dual tensor multiplet. This has an $SO(1,1)$ scalar manifold.  The  $\mathcal{N}=(2,0)$ supergravity theory considered in \cite{Samtleben:2019zrh}  involves adding four more self-dual tensors to complete the $(2,0)$ graviton multiplet, as well as adding scalars to extend the anti-self-dual tensor multiplet to $(2,0)$ supersymmetry. This  extends the $SO(1,1)$ to $SO(4+1,1)$. Superstrata go in the opposite sense, in that one adds $(1,0)$ anti-self-dual tensor multiplets, which extend  the $SO(1,1)$ to $SO(1,1+n)$, where $n$ is the number of added  anti-self-dual tensor multiplets.
 
From the three-dimensional perspective, these two extensions of the six-dimensional theory involves extending the $SO(4,4)$ of the consistent truncation of the basic theory  \cite{Cvetic:2000dm,Cvetic:2000zu,Cvetic:2000ah,Deger:2014ofa}  to $SO(4+4,4)$ or to $SO(4,4+n)$.  The parameter, $\varepsilon$ thus directly encodes whether we are adding self-dual or anti-self-dual tensors to the basic theory.  

As we noted earlier, flipping sign of $\varepsilon$ and $\alpha$ along with a change of orientation, leaves the  three-dimensional bosonic equations of motion unchanged in three dimensions.  The same is also true for the six-dimensional equations of motion: such an orientation flip on the three-dimensional base only changes the  duality conditions (\ref{eq:6Dselfduality}) in the six-dimensional bosonic equations of motion.  Thus, even in the six-dimensional theory, an $SO(p,q)$ theory with $p$ self-dual  and $q$ anti-self-dual multiplets, and the theory with $p$ and $q$ interchanged have exactly the same bosonic equations of motion (modulo orientations).

In  six-dimensional supergravity theories, there is  a correlation between the chirality of the supersymmetry and duality of the tensor gauge fields that belong to matter multiplets, or to the graviton multiplet.  The convention that is  used for superstrata, and is used in \cite{Samtleben:2019zrh},  is that the self-dual tensors belong to the  graviton multiplet and anti-self-dual tensors belong to matter multiplets.   If  one performed such an orientation flip on the $(2,0)$ theory, it would break the supersymmetry (to $(1,0)$) unless one also flipped the helicity of the spinors.  However, this is irrelevant to our goals here. We are focussed on the consistent truncation of the equations of motion. We will return to the supersymmetry in future work. 

Here, the important point is that the extensive work on consistent truncations to three dimensions that culminated in \cite{Samtleben:2019zrh}, shows that solutions to the three-dimensional equations of motion for the  $SO(4+4,4)$ theory, necessarily provide solutions to the equations of motion to the six-dimensional theory with five self-dual tensors multiplets and one anti-self-dual tensor multiplet.  A trivial orientation flip, means that this result maps onto the  $SO(4,4+4)$ theory in three dimensions and to a six-dimensional theory with one self-dual tensor multiplet and five anti-self-dual tensor multiplets.  In particular, this theory can be truncated to the $SO(4,4+1)$ theory and to the six-dimensional theory with one self-dual tensor multiplet two anti-self dual tensor multiplets\footnote{This truncation is trivially achieved by by imposing an invariance under the $SO(3)$  that acts on three of the five anti-self-dual tensor multiplets.}.

Thus the existence of the consistent truncation we seek is already guaranteed by the results of \cite{Hohm:2017wtr,Samtleben:2019zrh}.  What remains is to adapt the uplift formulae of \cite{Samtleben:2019zrh} to the theory of interest to us.   We will also subject our truncation and uplift formulae to  extensive and rigorous testing.

\subsection{The six-dimensional theory}
\label{sec:6Dtheorysub}

The general $\mathcal{N}=(1,0)$ six-dimensional supergravity theory coupled to an arbitrary number of tensor multiplets is discussed at length in \cite{Ferrara:1997gh,Riccioni:2001bg}. We will consider the $\mathcal{N}=(1,0)$ supergravity multiplet  coupled to two anti-self-dual tensor multiplets, as this is the relevant sector that captures the D1-D5-P solutions when reducing from ten-dimensional type IIB theory on a $\IT^4$ \cite{Giusto:2013rxa} (see also \cite{Duff:1998cr,Lavrinenko:1998hf}; or Appendix B of \cite{deLange:2015gca} for a quick summary).

\subsubsection{The six-dimensional theory for superstrata}
\label{sec:6Dtheory}

The six-dimensional bosonic field content we consider consists of  the metric $g_{\hmu\hnu}$, two scalars $\varphi,X$, and three three-forms $G^\hI$, $\hI=1,2,4$.\footnote{We conform to the idiosyncratic notation and conventions for the six-dimensional three-forms that is used in the superstrata literature. This slightly odd notation of omitting the index 3 is historical. In reduction of the six-dimensional system to five dimensions the $F$ and $d\beta$ fields are identified with $Z_{3}$ and $\Theta_{3}$ respectively.} The theory has a $SO(1,2)$ global symmetry, where the two scalars parametrize a $SO(1,2)/SO(2)$ coset.  The three-forms satisfy a self-duality relation:
\be \label{eq:6Dselfduality} \hat{*}G^I =  \frac{1}{\left(3!\right)^2} \epsilon\ind{_{\hmu\hnu\hrho}^{\hat{\alpha}\hat{\beta}\hat{\gamma}}} G^\hI_{\hat{\alpha}\hat{\beta}\hat{\gamma}} dx^\hmu\wedge dx^\hnu\wedge dx^\hrho = \varepsilon\, \mathcal{M}\ind{^\hI_\hJ} G^\hJ ,\ee
where $\varepsilon=\pm 1$, which serves as their equations of motion, together with their Bianchi identities:
\be \label{eq:6Dbianchi} dG^\hI = 0.\ee
We have parametrized the scalar self-duality matrix $\mathcal{M}$ as:
\be \cM\ind{^\hI_\hJ} = \frac{1}{2}\, e^{\sqrt{2}\varphi}\,\left( \begin{array}{ccc} X^2 & 8 & -2\sqrt{2} X\\ \frac18e^{-2\sqrt{2}\varphi} (2 +  e^{\sqrt{2}\varphi} X^2)^2 & X^2 & -\frac{X}{2\sqrt{2}}(2e^{-\sqrt{2}\varphi}+X^2) \\ +\frac{X}{\sqrt{2}}(2e^{-\sqrt{2}\varphi}+X^2) & 4\sqrt{2}X & -2e^{-\sqrt{2}\varphi} -2 X^2\end{array}\right), \label{Mdef}  \ee
Note that we use the following conventions for the $SO(1,2)$ metric which is used to raise or lower indices:
\be \eta^{\hI\hJ} = \left( \begin{array}{ccc} 0 & 1 & 0\\ 1 & 0 & 0\\ 0 & 0 & -2\end{array}\right). \label{etaDef}\ee
 The other bosonic equations of motion can be obtained by varying the pseudo-Lagrangian \cite{Ferrara:1997gh,Riccioni:2001bg}:
\be \label{eq:6Dlagr} \mathcal{L}_{6D} = R -\frac12 (\partial_\hmu\varphi)^2 -\frac12 e^{\sqrt{2}\varphi}(\partial_\hmu X)^2 - \frac16 \mathcal{M}_{\hI\hJ}G\ind{^\hI_{\hmu\hnu\hrho}}G^{\hJ\hmu\hnu\hrho}.\ee
Note that the scalar matrix $\mathcal{M}_{\hI\hJ}$ (with both indices down) is symmetric.

One should also note that the matrix $\tensor{\cM}{^\hI_\hJ}$ has one positive eigenvalue and two negative eigenvalues. It then follows from (\ref{eq:6Dselfduality}) that for superstrata (with two anti-self dual tensors)  one should take $\varepsilon=+1$, while for $\varepsilon=-1$, the theory has two self-dual tensors and hence the uplift formulae should reduce to a truncation of that given in \cite{Samtleben:2019zrh}.\footnote{Our conventions for the six-dimensional Hodge dual are given explicitly in (\ref{eq:6Dselfduality}). While never explicitly mentioned in \cite{Samtleben:2019zrh}, their convention for Hodge duals is such that their self-duality relation receives a relative minus sign compared to ours in (\ref{eq:6Dselfduality}).}

\subsection{The full six-dimensional uplift}
\label{sec:full6Duplift}

Here, we give the full ansatz for the non-linear KK reduction of the six-dimensional theory (\ref{eq:6Dlagr}) on an $S^3$, which gives the three-dimensional gauged supergravity discussed in Section \ref{Sect:3Dsugr}, with three-dimensional metric $g_{\mu\nu}$ (coordinates $x^\mu$), 14 scalars which consist of the four scalars $\chi_A$ and the 10 scalars parametrizing the symmetric matrix $m_{AB}$, and the six three-dimensional gauge fields parametrized by the antisymmetric $\tensor{\widetilde{A}}{_{\mu}^{AB}}$. This reduction ansatz follows from a simple adjustment of the ansatz considered in \cite{Samtleben:2019zrh} for six-dimensional $\mathcal{N}=(2,0)$ supergravity.

\subsubsection{The metric and scalars}

The six-dimensional metric ansatz is: 
\be
 \label{eq:6Dmetricansatz} ds_6^2 = (\det m_{AB})^{-1/2} \Delta^{1/2} ds_3^2  + g_0^{-2}(\det m_{AB})^{1/2} \Delta^{-1/2} m^{AB} \cD\mu^A \cD\mu^B,
\ee
where we have made the convenient re-definition of the three-dimensional gauge coupling:
\be \label{eq:g0gamma0} g_0 = 2 \gamma_0.\ee
We have also defined:
\be  \Delta = m_{AB}\mu^A \mu^B.
\label{eq:DeltaDef}
\ee
The four Cartesian coordinates, $\mu^A$, on $\IR^4$ are required satisfy $\mu^A\mu^A=1$, so as to define a unit $S^3$.  Their gauge-covariant derivatives are $\cD\mu^A = d\mu^A -2g_0 \tilde{A}^{AB}\mu^B$; see Appendix \ref{sec:app:S3} for more details. 
Note that the metric ansatz only depends on the scalar matrix $m_{AB}$ (and its inverse $m^{AB}$) and not on the scalars $\chi_A$. 
The six-dimensional scalars are given by the simple expressions:
\be \label{eq:6Dscalaransatz} e^{-\sqrt{2}\varphi} =  \Delta, \qquad X = \chi_A \mu^A .\ee

\subsubsection{The tensor gauge fields}

The expressions for the three-forms are quite unwieldy. It is easiest to give the two-form potentials $B^\hI$, related to the three-forms in the usual way:
\be G^\hI = dB^\hI.\ee
The three-forms $G^\hI$ and its two-form potentials $B^\hI$ can be decomposed as:\footnote{Our six-dimensional Hodge dual conventions are given explicitly in (\ref{eq:6Dselfduality}); for completeness, note that we take the six-dimensional Levi-Civita tensor to decompose as $\epsilon_{\mu\nu\rho ijk} = + \epsilon_{\mu\nu\rho} \epsilon_{ijk}$.}
\begin{align} G^\hI &= \frac{1}{3!} G^{\hI}_{ijk}\cD y^i\wedge \cD y^j\wedge \cD y^k + \frac12 G^\hI_{ij\mu}\cD y^i\wedge \cD y^j\wedge dx^\mu\\
\nonumber & + \frac12 G^\hI_{i\mu\nu}\cD y^i\wedge dx^\mu\wedge dx^\nu + \frac{1}{3!} G^\hI_{\mu\nu\rho}dx^\mu\wedge dx^\nu\wedge dx^\rho,\\
   B^\hI &= \frac12 B^\hI_{ij} \cD y^i\wedge \cD y^j + B^\hI_{i\mu} \cD y^i\wedge dx^\mu + \frac12 B^\hI_{\mu\nu} dx^\mu\wedge dx^\nu .
\end{align}

We will only give expressions for $B^\hI_{ij}$ and $B^\hI_{i\mu}$, which unambiguously determine the components $G^\hI_{ijk},G^\hI_{ij\mu}$ of the three-forms; the other components $G^\hI_{i\mu\nu},G^\hI_{\mu\nu\rho}$ (and thus also, by integration, $B^\hI_{\mu\nu}$) are then determined by the self-duality relation (\ref{eq:6Dselfduality}). The ansatze for $B^\hI_{ij}$ is (using the round sphere quantities defined in Appendix \ref{sec:app:S3}):
\begin{align}
 \label{eq:6DansatzB1ij} B^1_{ij} &= \left(-\frac{1}{g_0^2}\right)\left(-2\mathring{\omega}_{ijk}\mathring{\zeta}^k   + \frac12\mathring{\omega}_{ijk}\mathring{g}^{kl}\Delta\partial_l \left[\Delta^{-1}\right]     \right), \\
\label{eq:6DansatzB2ij}  B^2_{ij} &= \left(-\frac{1}{4g_0^2}\right)\left(4\, \varepsilon\, g_0^{-1}\alpha\,\mathring{\omega}_{ijk}\mathring{\zeta}^k  + \frac14\mathring{\omega}_{ijk}\mathring{g}^{kl}\Delta\partial_l \left[\Delta^{-1}X^2\right]    \right),\\
\label{eq:6DansatzB4ij}   B^4_{ij} &=\left(-\frac{\sqrt{2}}{g_0^2}\right) \frac12 \mathring{\omega}_{ijk}\mathring{g}^{kl}\Delta^{1/2}\partial_l \left(\Delta^{-1/2} X \right) .
\end{align}
while the ansatze for the components $B^\hI_{i\mu}$ is:
\begin{align}
\label{eq:6DansatzB1imu} B^1_{i\mu} &=  \left(-\frac{1}{g_0^2}\right)\partial_i\mu^A \left[2g_0\right] A^{AB}_\mu\left(\mu^B - 2\mathring{\zeta}^k\partial_k\mu^B\right), \\
 \label{eq:6DansatzB2imu} B^2_{i\mu} &= \left(-\frac{1}{4g_0^2}\right)\partial_i\mu^A\left[2g_0\right]\varepsilon \left( -\left[ \tensor{A}{_{\mu}^{A}_{B}} -2g_0^{-1}\alpha A_\mu^{AB}\right]\mu^B - 2g_0^{-1}\alpha A^{AB}_\mu\left[\mu^B- 2\mathring{\zeta}^k\partial_k\mu^B\right]\right),\\
  \label{eq:6DansatzB4imu} B^4_{i\mu} &= 0 .
\end{align}
In (\ref{eq:6DansatzB2imu}), the auxiliary gauge field $A\ind{^A_B}$ features, but can be integrated out in favor of the fundamental fields in three-dimensions using (\ref{Beqn}), (\ref{Bvecden}), and (\ref{dualGFsfinal}):
\be g_0 \tensor{A}{_{\mu}^{A}_{B}} -2 \alpha A^{AB}_\mu =- \frac{\varepsilon}{16}\epsilon_{ABCD}Y^{CD}_\mu +\frac{1}{8} \epsilon_{ABCD} \epsilon\ind{_\mu^{\nu\rho}}m_{CC'}m_{DD'} F^{C'D'}_{\nu\rho}.\ee
In Section \ref{sec:U1sqtrunc}, we will give explicit formulae for the entire three-forms $G^\hI$ in a specific sub-sector relevant for the $(1,0,n)$ superstrata.

The complete reduction ansatz is thus given by the metric ansatz (\ref{eq:6Dmetricansatz}), the scalar ansatz (\ref{eq:6Dscalaransatz}), and the two-form potential ansatze (\ref{eq:6DansatzB1ij})-(\ref{eq:6DansatzB4imu}). Note that there are two constant parameters $g_0$ and $\alpha$ in the uplift; these (or more precisely, their absolute value) can essentially be chosen at will, as there are two rescalings that one can perform on any six-dimensional solution which correspond to rescaling a three-dimensional theory; we discuss these in Appendix \ref{app:rescalings}. For example, as we will do in Section \ref{subsect:(1,m,n) 3D data}, a natural choice would be to choose $g_0$ such that $m_{AB}=\oneone$ at an asymptotic $AdS_3$ boundary; then $g_0^{-1}$ is identified with the (asymptotic) $S^3$ radius in the six-dimensional solution. The sign of $\alpha$ can be changed by changing $\varepsilon$, as discussed in Section \ref{sec:3Dsummary}.

Our reduction presented here is a simple modification and extension of the $\mathcal{N}=(2,0)$ $S^3$ reduction ansatz in \cite{Samtleben:2019zrh}; to match their reduction ansatz, we need to take $\varepsilon = -1$ and  the $SO(4)$ vector of three-forms $G^\alpha$ in  \cite{Samtleben:2019zrh} truncates to our three-form $G^4$ as $G^\alpha \sim \delta_{\alpha,1}G^4$, and accordingly for the three-dimensional scalars.  Appendix \ref{sec:app:Henning} contains the explicit matching of our results to those of in \cite{Samtleben:2019zrh}.  (This matching also involves some minor corrections to the uplift formulae presented \cite{Samtleben:2019zrh}.)

\subsubsection{Testing the consistent truncation}
\label{ss:testing}

As we noted earlier, the consistent truncation we are interested in is closely related to that of the $\mathcal{N}=(2,0)$ theory reduced on $S^3$.  Indeed, our observations in Section \ref{sec:Establishing} mean that our consistent truncation is essentially guaranteed.  However, we still need to establish our uplift formulae and ensure that we have all the details correct.  Our tests will also provide extensive and rigorous testing of the entire consistent truncation more broadly.
 
The first test will be to reduce the theory to a  $U(1)^2$ truncation.  Specifically, by imposing that the fields are invariant under a particular $U(1)$, we truncate the theory from an $SO(4)$ gauged theory to a $U(1)^2$ gauge theory. This is presented in Section \ref{sec:U1sqtrunc}.  For this reduced system  we explicitly checked  that the three-dimensional equation of motion (following from the truncated three-dimensional Lagrangian (\ref{eq:lagr3DU1sq})), together with our uplift ansatz to six-dimensions, imply the six-dimensional three-form Bianchi identities (\ref{eq:6Dbianchi}) and self-duality relations (\ref{eq:6Dselfduality}), as well as the six-dimensional scalar equations of motion (i.e. the equations of motion for $X,\varphi$ coming from (\ref{eq:6Dlagr})).

The second test involved constructing a new family of six-dimensional solutions:  the $(1,m,n)$ superstrata, which depend on two independent, arbitrary holomorphic functions of one variable.  We present these new six-dimensional solutions in Section \ref{Sect:superstrata} (and Appendix \ref{sect:6Dsuperstrata}) and show that  they precisely conform to our reduction ansatz.  We then extract the three-dimensional data in Section \ref{subsect:(1,m,n) 3D data}  and use this  as a detailed test of  the three-dimensional equations of motion.  

Finally,  in Section \ref{subsect:(1,0,n) 3D data}  we examine the overlap of our two tests by looking at the six-dimensional $(1,0,n)$ superstrata considered as part of the  $U(1)^2$ truncation of Section \ref{sec:U1sqtrunc}.

Needless to say, our uplift formulae, and the three-dimensional action, pass all of these tests.  More to the point, these tests provide multiple, independent cross checks of all the functional forms and their coefficients in all of our uplift formulae.  In particular, in the uplift we have thoroughly tested all the signs and numerical factors, as well as the appearances of $\alpha$, $g_0=2\gamma_0$ and $\varepsilon$, which correspond precisely to the parameters of the three-dimensional theory.

\subsection{\texorpdfstring{A $U(1)^2$}{U(1)2} truncation}\label{sec:U1sqtrunc}

In this section, we focus on a consistent truncation of the general reduction given in Section \ref{sec:full6Duplift}.  This truncation is most simply defined by restricting to the fields that are invariant under the $O(2) = U(1)$ subgroup of the $SO(4)$ gauge group that rotates the gauge indices $A=3,4$ into each other.   This truncation is the minimal one in which the $(1,0,n)$ superstrata sit, and provides an explicit, more approachable example of the complicated reduction formulae of Section \ref{sec:full6Duplift}.  There is an analogous truncation that restricts to the sector that is invariant under rotations that map  the gauge indices $A=1,2$ into each other; this is the minimal truncation that contains the $(1,1,n)$ superstrata.  Here we will focus on the first truncation.

We will use the explicit coordinates $(\theta,\varphi_1,\varphi_2)$ on the $S^3$, see (\ref{eq:appS3:explicitmu}). The $O(2)$ invariant gauge fields are simply the $U(1)^2$ Cartan sub-sector:
\begin{align}
A\ind{_\mu^{AB}} &= -\frac12 \left(\begin{array}{cc} \mathcal{A}^{\varphi_2}_{\mu} & 0_{2 \times 2}\\ 0_{2\times 2} & \mathcal{A}^{\varphi_1}_{\mu}\end{array}\right), & \mathcal{A}^{\varphi_i}_{\mu}&= \left(\begin{array}{cc} 0 & A^{\varphi_i}_{\mu}\\ -A^{\varphi_i}_{\mu} & 0 \end{array}\right),
\end{align}
which have been parametrized such that the resulting gauge-covariant coordinates are given by (see (\ref{eq:appS3:Dmu}) or (\ref{eq:appS3:Dy})):
 \be \cD\theta = d\theta, \quad \cD\varphi_i = d\varphi_i + g_0A^{\varphi_i} .\ee

Invariance under $O(2)$  means that  we  keep $\chi_1,\chi_2$ but set:
\be \chi_3 = \chi_4 = 0. \label{chiRestriction}\ee
It also  truncates the  10 scalars of $m_{AB}$ to 4 scalars $\xi_1,\xi_2,\xi_3,\xi_4$ defined as follows:
\be m_{AB} = \left( \begin{array}{cc} e^{-\xi_2} R & 0_{2 \times 2}\\ 0_{2 \times 2} & e^{-\xi_1}\oneone_{2 \times 2}\end{array}\right) , \qquad R = \exp\left(\xi_3 \left(\begin{array}{cc} \sin\xi_4 & \cos\xi_4\\ \cos\xi_4 & -\sin\xi_4\end{array}\right)\right)\,.\ee
Note that the gauge covariant derivatives on the scalars are given by:
\begin{align}
 \cD_\mu\xi_{1,2,3} &= \partial_\mu \xi_{1,2,3}, & \cD_\mu\xi_4 &= \partial_\mu\xi_4 +2g_0 A_\mu^{\varphi_1}  \\
 \cD_\mu\chi_1 &= \partial_\mu \chi_1 +g_0 A^{\varphi_1}_\mu \chi_2, & \cD_\mu\chi_2 &= \partial_\mu \chi_2 -g_0 A^{\varphi_1}_\mu \chi_1. 
\end{align}

Thus, the three-dimensional fields in this truncation are the metric $g_{\mu\nu}$, the two (Abelian) gauge fields $A^{\varphi_i}_\mu$, and the six scalars $\xi_1,\xi_2,\xi_3,\xi_4,\chi_1,\chi_2$. The three-dimensional Lagrangian is given by the appropriate truncation of the full three-dimensional Lagrangian (\ref{action-final}) and can be written explicitly as:
\begin{align}
\label{eq:lagr3DU1sq}  4\mathcal{L}_{3D,U(1)^2} &= R - \frac12 (\partial_\mu \xi_1)^2 - \frac12 (\partial_\mu \xi_2)^2 - \frac12 (\partial_\mu \xi_3)^2 - \frac12 \sinh^2\xi_3 ( \cD_\mu\xi_4)^2\\
 \nonumber & - \frac14 e^{-2\xi_1} F_{\mu\nu}^{\varphi_1} F^{\varphi_1,\mu\nu} - \frac14 e^{-2\xi_2} F_{\mu\nu}^{\varphi_2} F^{\varphi_2,\mu\nu}- \frac12e^{\xi_2}  \left( \cosh \xi_3\left[(\cD_\mu\chi_1)^2+(\cD_\mu\chi_2)^2\right]\right.\\
 \nonumber & \left.- \sinh \xi_3  \left[\sin \xi_4\left((\cD_\mu\chi_1)^2-(\cD_\mu\chi_2)^2\right)  +2\cos \xi_4 \cD_\mu\chi_1 \cD^\mu\chi_2  \right] \right)\\
 \nonumber& +e^{-1}\epsilon^{\mu\nu\rho}\left(  2\alpha A^{\varphi_1}_\mu F^{\varphi_2}_{\nu\rho} + \frac14 \,\varepsilon\, F_{\mu\nu}^{\varphi_2} (\chi_2\cD_\rho \chi_1 - \chi_1 \cD_\rho \chi_2) \right)- V ,\\
 \nonumber  V &= -2g_0^2 e^{\xi_1}\left(2 e^{\xi_2}\cosh\xi_3-e^{\xi_1}\sinh^2\xi_3\right) +\frac{g_0^2}{2} e^{2 \xi_1+\xi_2} \left[ e^{\xi_2}\left(\frac12\,\varepsilon\,\chi_1^2+\frac12\,\varepsilon\,\chi_2^2+4g_0^{-1}\alpha\right)^2 \right.\\
 \nonumber & \left. +\cosh \xi_3 \left(\chi_1^2+\chi_2^2\right)+\sinh \xi_3\left( (\chi_1^2-\chi_2^2)\sin \xi_4 +2 \chi_1 \chi_2 \cos \xi_4\right)  \right]
\end{align}
Note that simply $F^{\varphi_i} = dA^{\varphi_i}$.

We have checked that the three-dimensional equations of motion following from (\ref{eq:lagr3DU1sq}) and the reduction ansatz given in Section \ref{sec:full6Duplift} imply the six-dimensional three-form Bianchi identities and self-duality relations, as well as the six-dimensional scalar equations of motion.

The $(1,0,n)$ superstrata solution sits in this truncation (see below in Section \ref{subsect:(1,0,n) 3D data}).  It is a solution of (\ref{eq:lagr3DU1sq}) with $\xi_{3,4}=0$. Although one should note that setting these scalar fields to 0 does not give a consistent truncation of (\ref{eq:lagr3DU1sq}), many of the reduction formulae of Section \ref{sec:full6Duplift} simplify considerably when these scalars vanish.  First of all, we have:
\be \left.\Delta\right|_{U(1)^2, \xi_{3,4}=0} = m_{AB}\mu^A \mu^B =  e^{-\xi_2}\sin^2\theta + e^{-\xi_1}\cos^2\theta, \quad X = \sin\theta(\chi_1\sin\varphi_1+\chi_2\cos\varphi_1).\ee
The six-dimensional metric ansatz simplifies to:
\be \left.ds_6^2\right|_{U(1)^2, \xi_{3,4}=0} = e^{\xi_1+\xi_2}\Delta^{1/2}ds_3^2 + g_0^{-2} \left[ \Delta^{1/2}d\theta^2 + \Delta^{-1/2}e^{-\xi_1}\sin^2\theta \cD\varphi_1^2 + \Delta^{-1/2}e^{-\xi_2}\cos^2\theta \cD\varphi_2^2\right].\ee
In fact, it was this simple metric structure  that  led to the original conjecture  \cite{Bena:2017upb} that the $(1,0,n)$ superstrata should be part of a consistent truncation. 

Finally, we can also explicitly calculate all components of the three-forms, including those determined by self-duality, which we give here (for $\xi_{3,4}=0$) in form notation:\footnote{Note that all Hodge stars in (\ref{eq:U1sqG1})-(\ref{eq:U1sqG4}) refer to three-dimensional Hodge stars with metric $ds_3^2$; and $vol_3 = *1$ is the three-dimensional volume form.}
\begin{align}
\label{eq:U1sqG1} \left.G^1\right|_{U(1)^2,\xi_{3,4}=0} &= 2g_0^{-2}\Delta^{-2}e^{-\xi_1-\xi_2}\sin\theta\cos\theta d\theta\wedge \cD\varphi_1\wedge \cD\varphi_2\\
\nonumber & + g_0^{-2}\Delta^{-2}e^{-\xi_1-\xi_2}\sin^2\theta\cos^2\theta(d\xi_1-d\xi_2)\wedge \cD\varphi_1\wedge \cD\varphi_2\\
\nonumber &- g_0^{-1}\Delta^{-1}\left(e^{-\xi_2}\sin^2\theta F^{\varphi_2}\wedge \cD\varphi_1 + e^{-\xi_1}\cos^2\theta F^{\varphi_1}\wedge \cD\varphi_2\right)\\
\nonumber &- \frac12 e^{2\xi_1+2\xi_2}\,\varepsilon\,\left( 8\,\varepsilon\,\alpha + g_0(\chi_1^2 + \chi_2^2)\right) vol_3  ,
\end{align}
\begin{align}
\label{eq:U1sqG2} \left.4G^2\right|_{U(1)^2,\xi_{3,4}=0}  &= -g_0^{-2}\left(4\,\varepsilon\,\alpha g_0^{-1}+\frac12(\chi_1^2+\chi_2^2)\right)\sin\theta\cos\theta d\theta\wedge \cD\varphi_1\wedge \cD\varphi_2\\
  \nonumber   & + g_0^{-2} e^{-\xi_1-\xi_2}X^2\left(\Delta^{-2}+e^{\xi_1}\Delta^{-1}\right)\sin\theta\cos\theta d\theta\wedge \cD\varphi_1\wedge \cD\varphi_2\\
\nonumber   & + g_0^{-2}\,\varepsilon\,\sin\theta\cos\theta(-e^{-2\xi_1}*F^{\varphi_1}\wedge d\theta\wedge \cD\varphi_1 + e^{-2\xi_2}*F^{\varphi_2}\wedge d\theta\wedge \cD\varphi_2)\\
 \nonumber  & +g_0^{-2}X \cos\theta (\cos\varphi_1 D\chi_1 -\sin\varphi_1 D\chi_2)\wedge d\theta\wedge \cD\varphi_2\\
 \nonumber  & +\frac12g_0^{-2}X^2\Delta^{-2}e^{-\xi_1-\xi_2} \sin^2\theta\cos^2\theta (d\xi_1-d\xi_2)\wedge \cD\varphi_1\wedge \cD\varphi_2\\
 \nonumber  & -g_0^{-2} X e^{-\xi_1}\Delta^{-1}\sin\theta\cos^2\theta (\sin\varphi_1 D\chi_1+\cos\varphi_1 D\chi_2)\wedge \cD\varphi_1\wedge \cD\varphi_2\\
 \nonumber  &-g_0^{-1}\,\varepsilon\,\cos\theta \left(\sin\theta(*d\xi_1-*d\xi_2)+e^{\xi_2} X (\sin\varphi_1*D\chi_1+\cos\varphi_1*D\chi_2)\right)\wedge d\theta\\
 \nonumber  & -\frac12 g_0^{-1}X^2\Delta^{-1} (e^{-\xi_1}\cos^2\theta F^{\varphi_1}\wedge \cD\varphi_2 + e^{-\xi_2}\sin^2\theta F^{\varphi_2}\wedge \cD\varphi_1)\\
 \nonumber  & +g_0^{-1}\,\varepsilon\, e^{\xi_2}X \sin\theta (\sin\varphi_1*D\chi_2-\cos\varphi_1 *D\chi_1)\wedge \cD\varphi_1\\
 \nonumber  & +\,\varepsilon\,\left( 2 e^{\xi_1+\xi_2}g_0 -e^{2\xi_1+\xi_2}X^2\left[g_0+ e^{\xi_2}\left(2\,\varepsilon\,\alpha + \frac14g_0(\chi_1^2+\chi_2^2)\right)\right] \right)vol_3,
 \end{align}
 \begin{align}
\label{eq:U1sqG4} \left.\sqrt{2}G^4\right|_{U(1)^2,\xi_{3,4}=0}  &= g_0^{-2}\Delta^{-2}X e^{-\xi_1-\xi_2}(2+e^{\xi_1}\Delta)\sin\theta\cos\theta d\theta\wedge \cD\varphi_1\wedge \cD\varphi_2\\
\nonumber & + g_0^{-2} \cos\theta (\cos\varphi_1 D\chi_1-\sin\varphi_1D\chi_2)\wedge d\theta\wedge \cD\varphi_2\\
\nonumber  & + g_0^{-2}\Delta^{-2}\left[e^{-\xi_1-\xi_2}\sin^2\theta\cos^2\theta X(d\xi_1-d\xi_2) \right. \\
\nonumber  & \left.- \sin\theta \cos^2\theta \Delta e^{-\xi_1}(\sin\varphi_1 D\chi_1 +\cos\varphi_1 D\chi_2) \right]\wedge \cD\varphi_1\wedge \cD\varphi_2\\
\nonumber  & -g_0^{-1}\,\varepsilon\,e^{\xi_2}\cos\theta (\sin\varphi_1 *D\chi_1+\cos\varphi_1 * D\chi_2)\wedge d\theta \\
\nonumber  &  +g_0^{-1}\,\varepsilon\, e^{\xi_2} \sin\theta (\sin\varphi_1*D\chi_2-\cos\varphi_1 *D\chi_1)\wedge \cD\varphi_1\\
 \nonumber & -g_0^{-1}\Delta^{-1}X\left( e^{-\xi_1}\cos^2\theta F^{\varphi_1}\wedge \cD\varphi_2 + e^{-\xi_2}\sin^2\theta F^{\varphi_2}\wedge \cD\varphi_1\right)\\
 \nonumber & - \frac12\,\varepsilon\, e^{2\xi_1+\xi_2}X \left(2g_0 + e^{\xi_2}(8\,\varepsilon\,\alpha + g_0 (\chi_1^2+\chi_2^2))\right) vol_3 .
\end{align}

\section{Superstrata in three dimensions}
\label{Sect:superstrata}

To further test our uplift formulae (as well as develop the theory of superstrata), we have constructed a novel family of multi-mode superstrata, which are solutions of the six-dimensional theory of \ref{sec:6Dtheory}. We have also verified that they conform to the uplift formula of Section \ref{sec:full6Duplift}, and that they give a solution of the three-dimensional theory given by the action (\ref{action-final}). In the language of \cite{Heidmann:2019xrd}, this family is produced by superimposing the $(1,m,n)$ single-mode superstrata with $ m\in \{0,1 \} $ and $n\in \mathbb{Z}^{+}$. Since these are the maximal ranges allowed\footnote{These modes are restricted by regularity or equivalently CFT considerations, see \cite{Shigemori:2020yuo} for a discussion.} for $m$ and $n$, we refer to this family of solutions as the $(1,m,n)$ multi-mode family.

Appendix \ref{sect:6Dsuperstrata} contains the full system of six-dimensional BPS equations used to construct these solutions, the solutions themselves, along with the regularity and asymptotic charge analysis. Here we give the truncated three-dimensional data, which solve the equations of motion for the action (\ref{action-final}). In addition, in Section \ref{subsect:(1,0,n) 3D data} we discuss the simpler $(1,0,n)$ sub-family, which fits in the simpler $U(1)^{2}$ sub-sector of the six-dimensional reduction given in Section \ref{sec:U1sqtrunc}. 

\subsection{The holomorphic functions}
\label{subsect:(1,m,n) Specifying Solutions}
We use the $S^{3}$ coordinates $(\theta,\varphi_{1},\varphi_{2})$, with metric (\ref{dsSphere}). The $S^{3}$ is fibred over a ``deformed" $AdS_{3}$, which we parametrize by $(u,v,r)$, where 
\begin{align}
u = \frac{1}{\sqrt{2}}(t-y)  \,, \qquad v = \frac{1}{\sqrt{2}}(t+y)  \,, \label{uvDef}
\end{align}
are light cone coordinates, $t$ is the conventional time (in three dimensions) and $y$ parametrizes the common D1-D5 circle direction with radius $R_{y}$. 

Following \cite{Heidmann:2019xrd}, we introduce the complex coordinate
\begin{align}
\xi ~\equiv~\frac{r}{\sqrt{r^2+ a^2}} \, e^{i \frac{\sqrt{2} v}{R_y} }\,. \label{xidef}
\end{align}
A specific $(1,m,n)$ multi-mode superstrata is then fixed by specifying the two holomorphic functions:
\begin{align}
F_{0}=\sum_{n=1}^{\infty}b_{n}\xi^{n} \qquad \text{and} \qquad F_{1}=\sum_{n=1}^{\infty}d_{n}\xi^{n}\,, \label{F0F1def}
\end{align}
where $(b_{n},d_{n})$ are real numbers. Regularity of the solutions requires the introduction of the constant 
\begin{align}
c^{2} = \sum_{n=1}^{\infty} \left( b_{n}^{2} +d_{n}^{2} \right) \,, \label{cDef}
\end{align}
with the constraint:
\begin{align*}
\frac{2Q_{1}Q_{5}}{R_{y}^{2}} = 2a^{2}+c^{2}\,.
\end{align*}
See Appendix \ref{SubSect:Regularity} for details.

The $(1,0,n)$ multi-mode superstrata are recovered by setting $F_{1} =0 $, and the $(1,1,n)$ multi-mode superstrata are recovered by setting $F_{0} =0$. (These two multi-mode sub-families were first discussed in \cite{Heidmann:2019xrd}.)

\subsection{The three-dimensional description of \texorpdfstring{$(1,m,n)$}{(1,m,n)} superstrata}
\label{subsect:(1,m,n) 3D data}
We use the freedom discussed in Appendix \ref{app:rescalings} to rescale the six-dimensional uplift formulae of Section \ref{sec:full6Duplift} (using $\Lambda_2 = 2\sqrt{Q_1/Q_5}$), and then we choose:
\be \label{eq:3Dconsts1mn} \alpha = - \frac12 \varepsilon\, g_0, \qquad g_0 = (Q_1 Q_5)^{-1/4}.\ee
We have chosen these constant so that $g_0^{-1}=(Q_1 Q_5)^{1/4}$ corresponds to the radius of the $S^3$ in six-dimensions at the asymptotic $AdS_3\times S^3$ boundary, as appropriate for a D1-D5-P superstrata.

It is convenient to introduce the quantities:
\begin{align}
S_{A} = - \frac{aR_{y}g_{0}^{2}}{2\sqrt{ 2(a^{2}+r^{2}) }}  \left( iF_{0} ,F_{0},-ie^{i \frac{\sqrt{2}}{R_{y}} v} F_{1}, e^{i\frac{\sqrt{2}}{R_{y}} v }F_{1}  \right) + c.c. \,. \label{Sdef}
\end{align}
The four scalars, $\chi_A$, are then given by\footnote{An interesting perspective can also be gained from introducing the complex combinations:
\begin{align*}
z_{1}= S_{1}+iS_{2} \qquad \text{and} \qquad z_{2}=S_{3}+iS_{4}\,
\end{align*}
which simplifies some of the following expressions since:
\begin{align*}
S_{1}^{2}+S_{2}^{2}=\abs{z_{1}}^{2}\,, \qquad S_{3}^{2}+S_{4}^{2}=\abs{z_{2}}^{2}\,, \qquad S_{1}S_{3}-S_{2}S_{4} = \Re \left\lbrace z_{1}z_{2} \right\rbrace \,, \qquad S_{1}S_{4}+S_{2}S_{3} = \Im \left\lbrace z_{1}z_{2} \right\rbrace\,.
\end{align*}
}

\begin{align}
\chi_{A} = 2 S_{A} \,, \label{chiDef}
\end{align}
and the ten scalars in $m_{AB}$ are:
\begin{align}
m_{AB} &= \mathbb{I}-\begin{pmatrix}
S_{1}^{2}+S_{2}^{2} & 0 &  S_{1}S_{3}-S_{2}S_{4}  &  S_{1}S_{4}+S_{2}S_{3}  \\
0 & S_{1}^{2}+S_{2}^{2} &  S_{1}S_{4}+S_{2}S_{3}  & S_{2}S_{4}-S_{1}S_{3} \\
S_{1}S_{3}-S_{2}S_{4} & S_{1}S_{4}+S_{2}S_{3} & S_{3}^{2}+S_{4}^{4} & 0 \\
 S_{1}S_{4}+S_{2}S_{3} &  S_{2}S_{4}-S_{1}S_{3} & 0& S_{3}^{2}+S_{4}^{4}  \,. \label{m1mnSol}
\end{pmatrix}\,,
\end{align}
The three-dimensional metric takes the form of an $\mathbb{R}^{1}$ fiber over a conformally rescaled two-dimensional K\"ahler manifold:
\begin{align}
ds_{3}^{2} =  \frac{R_{y}^{2}g_{0}^{2}}{2} \left[ \Omega^{2}\, ds_{2}^{2} - a^{4}g_{0}^{4}\left(du + dv +    \frac{\sqrt{2} }{a^{2}R_{y} g_{0}^{4}}\,   \mathscr{A}\right)^{2} \right] \,, \label{ds3}
\end{align}
where: 
\begin{equation}
ds_{2}^{2} = \frac{\abs{d\xi}^{2}}{\left(1-\abs{\xi}^{2} \right)^{2}}    \,, \qquad \Omega^{2}= \frac{2 }{R_{y}^{2}g_{0}^{4}} \left( 1- S_{A}S_{A} \right) \,, \qquad
\mathscr{A} = \frac{i}{2} \left( \frac{\xi \, d\bar{\xi} - \bar{\xi} \, d\xi}{1-\abs{\xi}^{2}} \right)  \,. 
\label{ds3Data} 
\end{equation}

This shows that the three-dimensional metric has the form of a non-trivial, warped time-fibration over a non-compact $\IC\IP^1$.  This structure is almost certainly a consequence of supersymmetry and is extremely reminiscent of the  structure used to find Gutowski-Reall black holes in AdS$_5$ \cite{Gutowski:2004ez, Gutowski:2004yv}.

Finally, the six vector fields $\tensor{\widetilde{A}}{_{\mu}^{AB}}$ read:
\begin{align}
\tensor{\widetilde{A}}{_{\mu}^{AB}} dx^{\mu} &=  \frac{1}{\sqrt{2}a^{2}R_{y}g_{0}} \left(C_{1}\tensor{\eta}{_{1}^{AB}} + C_{2}\tensor{\eta}{_{2}^{AB}} + C_{3}\tensor{\eta}{_{3}^{AB}} + \bar{C}_{3}\tensor{\bar{\eta}}{_{3}^{AB}}\right)\,, \label{Aform}
\end{align}
where 
\begin{align}
C_{1} &=  \left(S_{1}S_{3}-S_{2}S_{4}  \right)\, \mathbf{d} \,, \\
C_{2} &=  \left(S_{1}S_{4}+S_{2}S_{3}  \right)\, \mathbf{d}  \,, \\
C_{3} &=   \left( \frac{a^{2}}{2} \right)  \, dv  -\frac{1}{2}\left( S_{1}^{2}+S_{2}^{2} -S_{3}^{2} - S_{4}^{2} \right)\, \mathbf{d}  \,, \\
\bar{C}_{3} &=  -\left( \frac{a^{2}+2r^{2}}{2} \right) \, dv  + \left(1- \frac{S_{A}S_{A}}{2}  \right)\, \mathbf{d} \,, \label{C3} \\
\mathbf{d} &= \frac{1}{\Omega^{2}}\left[  a^{4}( du+dv) +\frac{2r^{2}}{R_{y}^{2}g_{0}^{4}} \,  dv \right] \,,
\end{align}
and we have introduced the antisymmetric $4\times 4$ 't Hooft matrices, which may be written:
\begin{align}
\tensor{\eta}{_{1}^{AB}}&= \begin{pmatrix}
0 & \sigma_{x} \\
-\sigma_{x} & 0
\end{pmatrix} \,,  & \tensor{\eta}{_{2}^{AB}} &= \begin{pmatrix}
0 & -\sigma_{z} \\
\sigma_{z} & 0
\end{pmatrix} \,, & \tensor{\eta}{_{3}^{AB}} &= \begin{pmatrix}
i\sigma_{y} & 0 \\ 0 & i\sigma_{y}
\end{pmatrix} \,, \\
\tensor{\bar{\eta}}{_{1}^{AB}}&= \begin{pmatrix}
0 & -i \sigma_{y} \\
i \sigma_{y} & 0
\end{pmatrix} \,,  & \tensor{\bar{\eta}}{_{2}^{AB}} &= \begin{pmatrix}
0 & -\mathbb{I} \\
\mathbb{I}  & 0
\end{pmatrix} \,, & \tensor{\bar{\eta}}{_{3}^{AB}} &= \begin{pmatrix}
i\sigma_{y} & 0 \\ 0 & -i\sigma_{y}
\end{pmatrix}\,.
\end{align}

Note that $\eta_j$ and $\bar \eta_j$ generate the commuting $SU(2)$ factors of $SO(4) = SU(2) \times SU(2)$.  In particular, this means that the gauge fields in (\ref{Aform}) define an $SU(2) \times U(1)$ gauge connection.

We have explicitly checked that the three-dimensional fields $(ds_{3}^{2},\chi_{A},m_{AB},\tensor{\widetilde{A}}{_{\mu}^{AB}})$ given by (\ref{ds3}), (\ref{chiDef}), (\ref{m1mnSol}), and (\ref{Aform}) with the three-dimensional constants fixed by (\ref{eq:3Dconsts1mn}) (and (\ref{eq:g0gamma0})) satisfy the three-dimensional equations of motion coming from the three-dimensional Lagrangian (\ref{action-final}) (these are given explicitly in Appendix \ref{App:3Deom}). Note that the orientation of the three-dimensional manifold is tied to the sign of $\alpha$ as we must choose:
\begin{equation}
e^{-1}\epsilon_{uvr} = -\varepsilon\,, 
\label{eq:3Dorientation}
\end{equation}
where $e \equiv \sqrt{|\det(g_{\mu\nu})|}$.

\subsection{The solutions in \texorpdfstring{$U(1)^2$}{U(1)2} truncations}
\label{subsect:(1,0,n) 3D data}
The $(1,m,n)$ multi-mode solution simplifies greatly when one sets either $F_{0}=0$, so that $S_{3}=S_{4}=0$, or $F_{1}=0$, so that $S_{1}=S_{2}=0$. These are the $(1,0,n)$ and $(1,1,n)$ multimode solutions respectively, both introduced and analyzed in \cite{Heidmann:2019xrd}. In each instance the expansion of $\tensor{\widetilde{A}}{_{\mu}^{AB}}$ in (\ref{Aform}) simplifies with $C_{1}=C_{2}=0$, implying the $\tensor{\widetilde{A}}{_{\mu}^{AB}}$ define a $U(1)\times U(1)$ gauge connection. 

The $(1,0,n)$ multi-mode family conforms exactly to the $U(1)^{2}$ truncation of Section (\ref{sec:U1sqtrunc}). Using the notation of that section, the reduction data $(ds_{3}^{2},\chi_{1},\chi_{2},\xi_{1},\xi_{2},\xi_{3},\xi_4,A^{\varphi_{1}}_{\mu},A_{\mu}^{\varphi_{2}})$, are given by:
\begin{align}
\chi_{1,2} = 2 S_{1,2}\,,
\end{align}
with
\begin{align}
S_{1} = - \frac{ia R_{y}g_{0}^{2}}{2\sqrt{2(a^{2}+r^{2})}} \left(  F_{0}-\bar{F}_{0}\right)      \qquad \text{and} \qquad S_{2}= - \frac{aR_{y}g_{0}^{2}}{2\sqrt{2(a^{2}+r^{2})}}\left(  F_{0} + \bar{F}_{0}\right)  \,.
\end{align}
The three dimensional metric, $ds_{3}^{2}$, again takes the form (\ref{ds3})-(\ref{ds3Data}) but with the altered
\begin{align}
\Omega^{2} = \frac{2}{R_{y}^{2}g_{0}^{4}} (1-S_{1}^{2}-S_{2}^{2})\,.
\end{align}
The remainder of the scalars read:
\begin{align}
\xi_{1} = \xi_{3}=\xi_{4}=0  \qquad \text{and} \qquad e^{-\xi_{2}} = \frac{1}{2}R_{y}^{2}g_{0}^{4} \Omega^{2} \,.
\end{align}
While the  vector fields reduce to:
\begin{align}
 A^{\varphi_{1}}_{\mu}\, dx^{\mu} &= -\frac{a^{2}R_{y}g_{0}^{3}}{\sqrt{2}}(du+dv)   \,, \\
 A^{\varphi_{2}}_{\mu}\, dx^{\mu} &=  \frac{\sqrt{2}}{R_{y}g_{0} \Omega^{2}} \left[a^{2}(du+dv) + \frac{2}{a^{2}R_{y}^{2}g_{0}^{4}}  \left( (a^{2}+r^{2})(S_{1}^{2}+S_{2}^{2})  -a^{2}\right)\, dv\right]\,.
\end{align}

As noted earlier, the $(1,1,n)$ multi-mode family is part of another $O(2)$-invariant truncation. This has $\chi_{1}=\chi_{2}=0$ and non-trivial $(\chi_{3},\chi_{4})$. This  truncated theory will involve a non-trivial gauge coupling in the $\varphi_{1}$ direction rather than the $\varphi_{2}$ direction. A priori, one might guess that the $(1,0,n)$ and $(1,1,n)$ families are related by a simple change of coordinates $\varphi_{1}\leftrightarrow \varphi_{2}$. However, we see from (\ref{C3}) that the two solutions will have distinct gauge field expansions, even after re-labeling $\varphi_{1}\leftrightarrow \varphi_{2}$. This is in agreement with the work of \cite{Walker:2019ntz}, where it was also shown that the $(1,0,n)$ and $(1,1,n)$ single mode solutions are only equivalent after a non-trivial spectral transformation and reduction to five dimensions. 

\section{Final comments}
\label{sec:Conclusions}

We have shown that the  three-dimensional $(0,2)$  gauged $SO(4)$ supergravity described in Section \ref{Sect:3Dsugr} is a consistent truncation of six-dimensional  $(1,0)$ supergravity coupled to two tensor multiplets.  We have also shown that this consistent truncation includes the newly-constructed family of $(1,m,n)$ superstrata, which involve momentum waves encoded in two freely-choosable holomorphic functions of one variable. 

This raises the question as to whether there are other consistent truncations that might encode yet more classes of microstate geometries.  The answer is almost certainly yes.  First, the results of  \cite{Nicolai:2003ux} suggest that there may well be consistent truncations that  encode  higher KK modes, and even entire towers of such modes.  These KK towers include the modes of at least one tensor gauge field and so it seems  likely that this work could be extended to the  tensor  gauge fields that one needs for superstrata.  

There are also indications that the five-dimensional geometries that can be obtained from compactifications of the $(2,1,n)$ superstrata \cite{Bena:2017geu,Bena:2017upb,Walker:2019ntz}, may also give rise to consistent truncations.  These would be  gauged supergravity theories  in three dimensions obtained from AdS$_3$ $\times S^2$ compactifications of $\cN=2$ supergravity, coupled to vector multiplets, in five dimensions.

It therefore seems that the consistent truncations  described here might be the tip of an iceberg: there are almost certainly extensive generalizations of our results.  

As described in the introduction, our primary interest in examining these consistent truncations is to provide a new tool for the study of microstate geometries.  In this paper, we have shown that the consistent truncation contains large and interesting families of BPS superstrata.  We plan to see if the three-dimensional approach will enable us to find some new, broader families of BPS microstate geometries.  

One of the remarkable things about the  six-dimensional BPS equations is that, after specifying a hyper-K\"{a}hler base, the remaining equations reduce to a linear system \cite{Bena:2011dd}.  {\it A priori},  it is not clear whether this simplification will be manifest in the three-dimensional BPS equations. Indeed, it seems  likely that linearity in three dimensions will only emerge  if one restricts the gauge fields, $\tensor{\widetilde{A}}{_{\mu}^{AB}}$, to an Abelian sub-sector. These gauge fields may also need to be locked onto the scalar fields in some manner.  An important question then becomes, to what extent one can unlock all the non-abelian gauge fields, while still being able to solve the  BPS system?  The end result may well be intrinsically non-linear. If  solutions can still be found, then their uplift to six dimensions might reveal new hyper-K\"{a}hler bases, which may give new and interesting microstate geometries. As mentioned in the introduction, there is evidence that such bases should exist, coming from perturbation theory, in both \cite{Ceplak:2018pws,Tyukov:2018ypq}.

Even more important is the possibility of constructing non-BPS microstate geometries.  Through a simple parity flip, one can convert BPS superstata into anti-BPS superstata (ones that preserve a different, complementary set of supersymmetries).  It is therefore possible to use the three-dimensional formulation to study non-BPS configurations that start from a combination of BPS  and anti-BPS momentum waves.  It should be  relatively straightforward to set up a three-dimensional initial value  problem that should produce such  non-BPS microstate geometries as the result of `scattering'  BPS and anti-BPS waves.  The extent to which this can be done analytically, or semi-analytically, is unclear, but it will certainly be possible to study this numerically.

In considering the outcome of such an approach to non-BPS solutions, it is important to remember the Faustian bargain of consistent truncations.  It is quite possible that the combining of BPS and non-BPS solutions in three dimensions will evolve, at late times, into a singular solution.  As we have seen in many examples of microstate geometries, the appearance of a singularity in supergravity does not invalidate the microstate geometry program, but usually indicates that one has suppressed degrees of freedom that are essential to resolving the singularity.  Thus the appearance of a singularity at late times may simply be the result of  limiting the degrees of freedom to a consistent truncation.   

Even if singularities do arise in such non-BPS solutions, there will still be invaluable information to be gleaned from the three-dimensional analysis.  One will see the early, time-dependent behavior and the radiation that comes from the scattering.  By using the uplifts one may also be able to determine which degrees of freedom will be needed to resolve any singular behaviour.

It is also possible that the microstate geometries created in this way will be smooth and robust and provide families of non-BPS microstate geometries for which the holographic dictionary is precisely known. 

\vspace{0.8cm}

\section*{Acknowledgments}
\vspace{-2mm}
NPW would like to thank Henning Samtleben for his patient and careful explanations of many aspects of his work on gauged supergravity in three dimensions. DRM is supported by the ERC Starting Grant 679278 Emergent-BH. The work of NPW and RAW was supported in part by ERC Grant number: 787320 - QBH Structure and by DOE grant DE-SC0011687. RAW is very grateful to the IPhT of CEA-Saclay for hospitality during this project, his research was also supported by a Chateaubriand Fellowship of the Office for Science Technology of the Embassy of France in the United States.

\appendix

\section{The Three-sphere}\label{sec:app:S3}

It is convenient to parametrize the unit radius, round $S^3$ with four restricted Cartesian coordinates $\mu^A$ of $\IR^4$ that satisfy $\mu^A\mu^A = 1$, or alternatively with three (unrestricted) coordinates $y^i$. The round, unit-sphere metric in coordinates $y^i$ is $\mathring{g}_{ij}$, with corresponding completely antisymmetric tensor $\mathring{\omega}_{ijk}$. Following \cite{Samtleben:2019zrh}, we also use a vector $\mathring{\zeta}^i$ with unit divergence,
\be \mathring{\nabla}_i \mathring{\zeta}^i = 1.\ee

The  gauge-covariant derivatives on the sphere are then:
\be \label{eq:appS3:Dmu}  \cD\mu^A = d\mu^A -2 g_0 \tilde{A}^{AB}\mu^B, \ee
where $g_0=2\gamma_0$ is the gauge coupling, and we have used the dual gauge fields given in (\ref{dualGFsfinal}). We can rewrite this as:
\be \cD\mu^A = \partial_i\mu^A \cD y^i,\ee
with:
\be \label{eq:appS3:Dy} \cD y^i = dy^i -2 g_0 \mathcal{K}^i_{AB} \tilde{A}^{AB} ,\ee
where we have used the Killing vectors on the sphere:
\be \mathcal{K}^i_{AB} = \mathring{g}^{ij}\partial_j \mu^{[A} \mu^{B]}.\ee
There are many other identities involving the $\mu^A$ (which we will not explicitly need in this paper); see, for example, Appendix A of \cite{Samtleben:2019zrh}.

An explicit coordinate basis that can be used is, for example, the standard coordinates $y^i=(\theta,\varphi_{1},\varphi_{2})$ with:
 \be \label{eq:appS3:explicitmu} \mu^1 = \sin\theta \sin\varphi_1, \quad \mu^2 = \sin\theta \cos\varphi_1, \quad \mu^3 = \cos\theta \sin\varphi_2, \quad \mu^4 = \cos\theta\cos\varphi_2.\ee
The metric in these coordinates of the unit radius round $S^3$ is:
\be \mathring{ds}^2_{S^3} = \mathring{g}_{ij}dy^idy^j = d\theta^2 + \sin^2\theta d\varphi_1^2 + \cos^2\theta d\varphi_2^2, \label{dsSphere}\ee
so that $\mathring{\omega}_{ijk} = (\sin\theta\cos\theta) \epsilon_{ijk}$, with $\epsilon_{123}=+1$ and completely antisymmetric. In these coordinates, we can take:
\be \mathring{\zeta}^i = \left( \frac12\tan\theta,0,0\right).\ee

\section{Six-dimensional and three-dimensional rescalings}
\label{app:rescalings}

We wish to point out two rescalings of the six-dimensional fields which have a counterpart as a rescaling of three-dimensional fields, through the uplift formulae in Section \ref{sec:full6Duplift}.

The first rescaling is:
\begin{align}
 e^{\sqrt{2}\varphi} & \rightarrow \Lambda_1^{-2}\, e^{\sqrt{2}\varphi} , & X & \rightarrow \Lambda_1\, X, &  g^{(6D)}_{\hmu\hnu} &\rightarrow \Lambda_1\, g^{(6D)}_{\hmu\hnu},\\
 G^1 &\rightarrow G^1, & G^2 &\rightarrow \Lambda_1^2\, G^2, & G^4 &\rightarrow \Lambda_1\, G^4,
\end{align}
which corresponds to the three-dimensional rescaling:
\begin{align}
 m_{AB} &\rightarrow \Lambda_1^2\, m_{AB}, & \chi_A & \rightarrow \Lambda_1\, \chi_A,\\
 \alpha &\rightarrow \Lambda_1^2\, \alpha, & g^{(3D)}_{\mu\nu} &\rightarrow \Lambda_1^4\, g_{\mu\nu}^{(3D)}.
\end{align}
Under this scaling, the six-dimensional Lagrangian (\ref{eq:6Dlagr}), resp. three-dimensional action (\ref{action-final}), scales as $\hat{e}\,\mathcal{L}_{6D}\rightarrow \Lambda_1^2\, \hat{e}\,\mathcal{L}_{6D}$, resp. $e\,\mathcal{L}_{3D}\rightarrow \Lambda_1^2\, e\, \mathcal{L}_{3D}$ (with $\hat{e} = \sqrt{-\det g_{\hmu\hnu}^{(6D)}}$ and $e=\sqrt{-\det g_{\mu\nu}^{(3D)}}$).

The second rescaling is:
\begin{align}
 e^{\sqrt{2}\varphi}  & \rightarrow \Lambda_2^{-2}\, e^{\sqrt{2}\varphi} , & X & \rightarrow \Lambda_2\, X, & g^{(6D)}_{\hmu\hnu} &\rightarrow  g^{(6D)}_{\hmu\hnu},\\
 G^1 &\rightarrow \Lambda_2^{-1}\,G^1, & G^2 &\rightarrow \Lambda_2\, G^2, & G^4 &\rightarrow G^4,
\end{align}
which has the three-dimensional counterpart:
\begin{align}
 m_{AB} &\rightarrow \Lambda_2^2\, m_{AB}, & \chi_A & \rightarrow \Lambda_2\, \chi_A,\\
 \alpha &\rightarrow \Lambda_2^{5/2}\, \alpha, & g_{\mu\nu}^{(3D)} &\rightarrow \Lambda_2^3\, g_{\mu\nu}^{(3D)},\\
 g_0 &\rightarrow \Lambda_2^{1/2}\,g_0, & A^{AB}_{\mu} &\rightarrow \Lambda_2^{-1/2}\, A^{AB}_\mu.
\end{align}
Note that the rescaling of $g_0$ implies the same rescaling of $\gamma_0$ through (\ref{eq:g0gamma0}).
The six-dimensional action (\ref{eq:6Dlagr}) is invariant under this scaling, $\hat{e}\,\mathcal{L}_{6D}\rightarrow \hat{e}\,\mathcal{L}_{6D}$, while the three-dimensional action (\ref{action-final}) rescales as $e\,\mathcal{L}_{3D}\rightarrow \Lambda_2^{3/2}\, e\,\mathcal{L}_{3D}$.

A combination of both of these scalings can be used to rescale the two constants $|\alpha|$ and $g_0$ to any value in the reduced three-dimensional theory.

\section{Matching with the conventions of \texorpdfstring{\cite{Samtleben:2019zrh}}{Samtleben-Sar{\i}oglu}}
\label{sec:app:Henning}

Our goal here is to provide a map between our conventions and those of \cite{Samtleben:2019zrh}.

\subsection{Consistency of gauge field actions, Chern-Simons term and conventions}
\label{sec:app:gauge}

There is some tension between our  formulation of the gauge action, and that of  \cite{Samtleben:2019zrh}.  Here we discuss the differences in detail and describe why  we have provided a consistent set of conventions.

Essentially there are four signs that must be correctly correlated:  (i) The sign of the representation matrices that define the minimal couplings, (ii) the sign of the structure constants,  (iii) the sign of the  $A \wedge A $ term in the field strength, and (iv) the sign of the $A \wedge A\wedge A$ term in the CS action.

We have defined the covariant derivative by (\ref{covderiv1}):
\begin{equation}
\widehat\cD_\mu \, \cX_{\bar P}  ~\equiv~ \partial_\mu \, \cX_{\bar P}  ~+~ {A_\mu}^{\bar K \bar L} \, \Theta_{\bar K \bar L, \bar M \bar N } \, (T^{\bar M \bar N})_{\bar P}{}^{\bar Q} (\cX_{\bar Q} )   \,, 
\label{covderiv1a}
\end{equation}
and  have chosen the representation matrices as so as to obtain (\ref{Tform}), which means we have taken:
\begin{equation}
 (T^{\bar M \bar N})_{\bar P}{}^{\bar Q}  ~=~ \delta^{\bar N}_{\bar P} \, \eta^{\bar M \bar Q} ~-~ \delta^{\bar M}_{\bar P}\, \eta^{\bar N \bar Q}   \,.
 \label{Tmatrix2}
\end{equation}
This then led to the covariant derivatives given in (\ref{covderiv2}) and ultimately to the covariant derivatives in (\ref{covderiv4}).
\begin{equation}
\begin{aligned}
\cD_\mu \, \cX_{A}  &~=~ \partial_\mu \, \cX_{A}   ~-~  4\, \gamma_0\,\widetilde A_\mu{}^{AB} \, \cX_{B} \,, \\  
 \cD_\mu \, \cX^{A}  &~=~ \partial_\mu \, \cX^{A}    ~-~  4\, \gamma_0\,\widetilde A_\mu{}^{AB} \, \cX^{B} \,, \qquad  \cD_\mu \, \cX_0 ~=~ \partial_\mu \, \cX_0  \,.
\end{aligned}
\label{covderiv4a}
\end{equation}
We note that it follows from this that the $SO(4)$ field strength is given by
\begin{equation}
\begin{aligned}
\big[\, \cD_\mu \, , \cD_\nu \,] \,\cX_{A}   & ~\equiv~ - 4\, \gamma_0\,   \widetilde F_{\mu \nu}{}^{AB} \,  \cX_{B} \\
   & ~=~ - 4\, \gamma_0\, \Big(\partial_\mu \widetilde A_\nu{}^{AB} - \partial_\nu \widetilde A_\mu{}^{AB} ~-~ 4\, \gamma_0 \,  \big (\widetilde A_\mu{}^{AC} \,  \widetilde A_\nu{}^{CB} -  \widetilde A_\nu{}^{AC} \,  \widetilde A_\mu{}^{CB} \big)\,\Big) \cX_{B}   \,,
 \end{aligned}
\label{fieldstrength1}
\end{equation}
from which one obtains
\begin{equation}
F_{\mu \nu}{}^{AB}  ~=~ \coeff{1}{2}\, \epsilon_{ABCD} \,  \widetilde F_{\mu \nu}{}^{CD}    ~=~ 2\,\big( \partial_{[\nu}   A_{\rho]}{}^{AB}   ~-~4 \,  \gamma_0\,   A_{[\nu} {}^{C[A} \,\widetilde  A_{\rho]} {}^{B]C}\big)  \,.
\label{fieldstrength2}
\end{equation}
It is necessary for consistency, that this is precisely the field strength given in (\ref{Maxdefn}).  The latter expression was obtained from the Chern-Simons action after integrating out the $\IT^6$ gauge fields, ${B_\mu}{}^{AB}$.  The important message here is that our explicit expressions for the gauge covariant derivatives and gauge actions are consistent with one another.

The simplest way to obtain the Chern-Simons action is to work with $F\wedge F$ in higher dimensions and write it as  $d(A \wedge F + \frac{1}{3} A \wedge A \wedge A)$, and then the term in parentheses is the action we seek.
This leads to 
\begin{equation}
F^a  ~=~  dA^a ~+~ \coeff{1}{2} \, f^a{}_{bc} \,A^b \wedge A^c \,, \qquad \cL_{CS}   ~=~  A^a \wedge dA^a ~+~ \coeff{1}{3} \, f^a{}_{bc} \,A^a\wedge A^b \wedge A^c \,,
\label{Maxterms}
\end{equation}
from which it follows that the Chern-Simons we seek is given by (\ref{CSgeneric}):
\begin{equation}
\cL_{CS} ~=~  \frac{1}{4}\, \varepsilon^{\mu \nu \rho} \,  A_\mu{}^{\bar K\bar L} \,\Theta_{\bar K\bar L,\bar M\bar N}\,
\Big( \partial_\nu A_\rho {}^{\bar M\bar N}  ~+~  \frac{1}{3}\, f^{\bar M\bar N,\bar P\bar Q}{}_{\bar R\bar S}\,\Theta_{\bar P\bar Q,\bar U\bar V}\,A_\nu {}^{\bar U\bar V}\, A_\rho{}^{\bar R\bar S} \Big)\,,
\label{CScorrected}
\end{equation}
In passing from the general expression (\ref{Maxterms}) to (\ref{CScorrected}), one should remember that the role of the embedding tensor, $\Theta$, is simply that of a projector from the large algebra, $SO(4,5)$, down to the gauge algebra, $SO(4) \times \IT^6$ and thus  (\ref{Maxterms}) is the appropriate expression on the gauge Lie algebra after projection.

The corresponding expression in   \cite{Samtleben:2019zrh} (equation (2.6)) is ``non-canonical" in that the sign of the $A \wedge A\wedge A$ term is reversed relative to our ``canonical choice.''  Earlier references, like \cite{Nicolai:2001sv, Nicolai:2001ac}, have the  canonical form of the Cherns-Simons term, (\ref{CScorrected}).

One should also note that our choice of representation matrices, (\ref{Tmatrix2}), leads to the opposite sign of the commutators and structure constants, (\ref{comms}) and (\ref{structureconsts}),  when compared to  \cite{Samtleben:2019zrh}.  This means that our final Chern-Simons action actually matches that of \cite{Samtleben:2019zrh}.  However, our minimal couplings and representation matrices have the opposite sign to those of \cite{Samtleben:2019zrh} and thus we believe there is a potential  inconsistency  in the complete action of \cite{Samtleben:2019zrh}.

We have used the ``canonical" form of the Chern-Simons terms, together with a consistent choice the generators (\ref{Tmatrix2}) and field strength.   The equations of motion are sensitive to all of these sign choices and the fact that the $(1,m,n)$ superstrata solve the resulting equations of motion, with all of these convention, give us further confidence that our conventions are consistent.

\subsection{Matching the uplift formulae}
\label{sec:app:uplift}

 We start by noting that our three-dimensional formulation matches that of \cite{Samtleben:2019zrh} if one sets
\begin{equation}
\label{eq:3Dmatch}  \gamma_0 ~=~  1 \qquad \Leftrightarrow \qquad g_0 ~=~  2 \  \,.
\end{equation}
The parameter $\alpha$ is the same in both sources.

Turning to the six-dimensional theory,  our scalar matrix is
\begin{equation}
 \cM^{\hI \hJ} ~=~ \frac{1}{2}\, e^{\sqrt{2}\varphi}\,\left( \begin{array}{ccc} 8 & X^2  &4 \sqrt{2} X\\ 
 X^2 &  \frac18e^{-2\sqrt{2}\varphi} (2 +  e^{\sqrt{2}\varphi} X^2)^2 &  \frac{1}{\sqrt{2}}(2e^{-\sqrt{2}\varphi}+X^2)X 
 \\   4\sqrt{2}X &\frac{1}{\sqrt{2}}(2e^{-\sqrt{2}\varphi}+X^2) X&4 \,(e^{-\sqrt{2}\varphi} + X^2) \end{array}\right) \,, \label{eq:MUs}   
\end{equation}
where we have raised the second index using our $SO(1,2)$ metric:
\begin{equation}
\eta^{\hI\hJ} ~=~ \left( \begin{array}{ccc} 0 & 1 & 0\\ 1 & 0 & 0\\ 0 & 0 & -2 \end{array}\right)\,.
 \label{eq:etaUs}
\end{equation}
One should also recall (\ref{eq:DeltaDef}) and  (\ref{eq:6Dscalaransatz}): 
\begin{equation}
e^{-\sqrt{2}\varphi}  ~=~ \Delta  ~=~ m_{AB}\mu^A \mu^B  \,.
\end{equation}

The corresponding objects in \cite{Samtleben:2019zrh} are 
\begin{equation} 
\begin{aligned}
 & \widetilde \cM^{a b} ~=~ \\
 & \frac{1}{8} \,\left( \begin{array}{ccc} 4 \Delta+ 4 X^2+ \Delta^{-1}(2+X^2)^2 &   4 \Delta^{-1} - \Delta (2+ \Delta^{-1} X^2)^2  & -2 \sqrt{2} (2 + \Delta^{-1}(2+X^2)) X\\ 
 4 \Delta^{-1} - \Delta (2+\Delta^{-1} X^2)^2 &  4 \Delta+ 4 X^2+ \Delta^{-1}(2-X^2)^2   &   2 \sqrt{2} (2 - \Delta^{-1}(2-X^2)) X
 \\   -2 \sqrt{2} (2 + \Delta^{-1}(2+X^2)) X  &2 \sqrt{2} (2 - \Delta^{-1}(2-X^2)) X &1+ \Delta^{-1} X^2 \end{array}\right) \,, 
 \end{aligned}
 \label{eq:MUSSs}   
\end{equation}
and 
\begin{equation}
\tilde \eta^{ab} ~=~ \left( \begin{array}{ccc} 1 & 0 & 0\\ 0 & -1 & 0\\ 0 & 0 & -1 \end{array}\right)\,,
 \label{eq:etaSS}
\end{equation}
where the indices $a,b,\ldots$ take the values $0, \bar 0, \bar1$. To compare with the results of \cite{Samtleben:2019zrh}, one should note that \cite{Samtleben:2019zrh} uses two expressions: $\Delta$ and $\tilde\Delta$. We re-label the $\Delta$ of \cite{Samtleben:2019zrh} as $\hat\Delta$ here; the relations between these quantities and our $\Delta$ are:
\begin{equation}
 \label{eq:phiDelta}
\hat \Delta ~=~  \big(\det(m_{AB})\big)^{\frac{1}{4}} \,  \Delta^{-\frac{1}{4}} \, , \qquad \tilde \Delta   ~=~ \Delta^{-\frac{1}{4}}    \,. 
\end{equation}

Define the matrix
\begin{equation}
P  ~=~ \left( \begin{array}{ccc} -\frac{1}{2\sqrt{2}}    & -  \sqrt{2}   &0 \\ 
 -\frac{1}{2\sqrt{2}} & \sqrt{2}   & 0
 \\  0 &0 &  \frac{1}{\sqrt{2}}   \end{array}\right) \,, \label{eq:Pdefn}   
\end{equation}
then one can easily verify that 
\begin{equation}
\tilde \eta  ~=~  P \, \eta \, P^t \,, \qquad \widetilde \cM   ~=~  P \, \cM \, P^t \,,  \label{eq:Props}   
\end{equation}
Thus $P$ provides a change of basis from our fields to those of \cite{Samtleben:2019zrh}.  In particular, performing the change of basis on the gauge potentials (\ref{eq:6DansatzB1ij})--(\ref{eq:6DansatzB4ij}) yields
\begin{align}
B^0_{ij}   ~=~   & -\frac{1}{2\sqrt{2}} \, B^1_{ij}  ~-~ \sqrt{2} \, B^2_{ij}  \nonumber \\
~=~ &\frac{1}{\sqrt{2} \,g_0^2} \,\Big[ -\big(1-2\,\varepsilon \alpha g_0^{-1}\big) \,\mathring{\omega}_{ijk}\mathring{\zeta}^k   -  \coeff{1}{4}\,\mathring{\omega}_{ijk}\mathring{g}^{kl}\partial_l (\log \Delta) ~+~ \coeff{1}{8} \, \mathring{\omega}_{ijk}\mathring{g}^{kl}\Delta\partial_l ( \Delta^{-1}X^2)  \Big]  \,, \label{eq:6DB0ij}    \\
B^{\bar 0}_{ij}   ~=~ & -\frac{1}{2\sqrt{2}} \, B^1_{ij}  ~+~ \sqrt{2} \, B^2_{ij} \nonumber \\
~=~ &\frac{1}{\sqrt{2} \,g_0^2} \,\Big[ -\big(1 + 2\,\varepsilon \alpha g_0^{-1}\big) \,\mathring{\omega}_{ijk}\mathring{\zeta}^k   -  \coeff{1}{4}\,\mathring{\omega}_{ijk}\mathring{g}^{kl}\partial_l (\log \Delta) ~-~ \coeff{1}{8} \, \mathring{\omega}_{ijk}\mathring{g}^{kl}\Delta\partial_l ( \Delta^{-1}X^2)  \Big] \,, \label{eq:6DB0barij}    \\
B^{\bar 1}_{ij}   ~=~ & \frac{1}{\sqrt{2}}   \, B^4_{ij} ~=~ - \frac{1}{2\,g_0^2} \,\mathring{\omega}_{ijk}\mathring{g}^{kl}\Delta^{1/2}\partial_l \left(\Delta^{-1/2} X \right) \,, \label{eq:6DB1barij}    
\end{align}
Similarly, transforming (\ref{eq:6DansatzB1imu}) -- (\ref{eq:6DansatzB2imu}) yields
\begin{align}
B^0_{\mu i}   ~=~ & -\frac{1}{2\sqrt{2}} \, B^1_{i \mu}  ~+~ \sqrt{2} \, B^2_{i \mu}  \nonumber \\
~=~ & -\frac{1}{\sqrt{2} \,g_0} \,\big( \partial_i\mu^A\big)\, \Big[  \big(A^{AB}_\mu  -  \varepsilon \, {A_\mu}^A{}_B \big) \mu^B ~-~2\, \big(1-2\,\varepsilon \alpha g_0^{-1}\big)\,A^{AB}_\mu \big(\mathring{\zeta}^k\partial_k\mu^B\big) \Big]  \,,  \label{eq:6DB0imu}    \\
B^{\bar 0}_{\mu i}   ~=~ &   \frac{1}{2\sqrt{2}} \, B^1_{i \mu}  ~-~ \sqrt{2} \, B^2_{i \mu}  \nonumber \\
~=~ & -\frac{1}{\sqrt{2} \,g_0} \,\big( \partial_i\mu^A\big)\, \Big[  \big(A^{AB}_\mu  + \varepsilon \, {A_\mu}^A{}_B \big) \mu^B ~-~2\, \big(1+ 2\,\varepsilon \alpha g_0^{-1}\big)\,A^{AB}_\mu \big(\mathring{\zeta}^k\partial_k\mu^B\big) \Big]\,,\label{eq:6DB0barimu}    \\
B^{\bar 1}_{ \mu i }   ~=~ & - \frac{1}{2\sqrt{2}}   \, B^4_{i \mu} ~=~  0 \,. \label{eq:6DB1barimu}    
\end{align}
Note that we have reversed the indices, ${ \mu i }$, on the left-hand side to facilitate comparison with \cite{Samtleben:2019zrh}.  

We find a perfect match for the $B_{ij}$ components, up to an overall factor of $g_0^{-2}$, provided that one uses  (\ref{eq:3Dmatch}) and takes
\begin{equation}
\varepsilon ~=~ -1     \,.  \label{eq:translation1}   
\end{equation}
This choice was anticipated in Sections \ref{sec:scalars} and  \ref{sec:6Dtheory}.

Using (\ref{eq:3Dmatch}) and (\ref{eq:translation1}) we also find  a nearly perfect match for the $B_{\mu i}$ components, up to an overall factor of $-g_0^{-1}$. 

The overall factors of $g_0^{-2}$ in $B_{ij}$  and $-g_0^{-1}$  in $B_{\mu i}$  are easily fixed using the scalings in Appendix \ref{app:rescalings}.  Indeed, choosing 
\begin{equation}
 \label{eq:scales} \Lambda_1  ~=~ \Lambda_2^{-1} ~=~ \Lambda  
\end{equation}
results in the six-dimensional rescaling 
\begin{equation}
g^{(6D)}_{\hmu\hnu} ~\rightarrow~  \Lambda \, g^{(6D)}_{\hmu\hnu} \,, \qquad G^\hI ~\rightarrow~ \Lambda \,  G^\hI  \,.
\end{equation}
One can then take $\Lambda =g_0^2 =4$ to match the overall  scale in $B_{ij}$.   

This scaling then creates an overall factor of $-g_0$ in $B_{\mu i}$.  This can then be compensated by coordinate changes:
\begin{equation}
x^\mu ~=~  \coeff{1}{2} \, \tilde x^\mu \,, \qquad   y^i ~=~ -  \tilde{y}^i  \,,
\end{equation}
which then rescale $B_{\mu i}$ by $-\frac{1}{2}$, while leaving $B_{ij}$ unchanged. 

There are two discrepancies between our analysis and that of \cite{Samtleben:2019zrh} that may be  transcription errors in \cite{Samtleben:2019zrh}. First, although our expressions for $B^{0}_{ij},B^{\bar 0}_{ij},B^{\bar 1}_{ij}$ match those in  \cite{Samtleben:2019zrh} precisely, our expressions for $B^{0}_{\mu i}$ and $B^{\bar 0}_{\mu i}$ match the expressions in \cite{Samtleben:2019zrh} for ${B _{\mu i}}{}_0$ and ${B_{\mu i}}{}_{\bar 0}$.  When $0$ and ${\bar 0}$ are lowered using $\tilde \eta_{ab}$ (see,  \ref{eq:etaSS}), they get a relative minus sign and so they cannot be reconciled simultaneously.  

Second, we have, using (\ref{eq:3Dmatch}):
\begin{equation}
\label{eq:appS3:Dya} \cD y^i ~=~ dy^i -2 g_0 \mathcal{K}^i_{AB} \tilde{A}^{AB} ~=~ -d\tilde y^i - 4\, \mathcal{K}^i_{AB} \tilde{A}^{AB}  \,,
\end{equation}
%
whereas, \cite{Samtleben:2019zrh}  defines
\begin{equation}
 \label{eq:appS3:Dyb} \cD y^i ~=~  dy^i  +  \mathcal{K}^i_{AB} \tilde{A}^{AB}\,,
\end{equation}
so these expressions do not match. We also note that the combination $g_0  A^{AB}_\mu dx^\mu$ is both coordinate invariant and invariant under both the rescalings described in Appendix \ref{app:rescalings}, and so we cannot reconcile our expressions for $\cD y^i $ with those of \cite{Samtleben:2019zrh}.

Therefore, up to two minor discrepancies, our results match  the expressions in \cite{Samtleben:2019zrh}.    As we indicated in  Section \ref{ss:testing}, we have subjected our uplift formulae to rigorous testing in both three and six dimensions, and have every confidence in our expressions and normalizations.

\section{\texorpdfstring{$(1,m,n)$}{(1,m,n)} Superstrata in six dimensions} \label{sect:6Dsuperstrata}
This appendix summarizes the six-dimensional BPS equations for the  D1-D5-P system, and the novel construction of the $(1,m,n)$ multi-mode superstrata family of solutions.

\subsection{Six-dimensional BPS equations}
\label{app:BPS equations}
All \nBPS{8} solutions of six-dimensional, $\mathcal{N}=(1,0)$  supergravity, coupled to 2 tensor multiplets, with the same charges as the D1-D5-P system, satisfy a ``layered" set of linear equations. These equations were first developed in \cite{Bena:2011dd} for a single tensor multiplet, and extended to include a second tensor multiplet in \cite{Giusto:2013rxa}. We follow \cite{Bena:2017geu}, introducing the equations in an explicitly $SO(1,2)$ covariant form, using the $SO(1,2)$ indices: $\hat{I},\hat{J},\hat{K},\cdots \in \{1,2,4 \}$, with non-zero $SO(1,2)$ metric components (see (\ref{etaDef})):
\begin{align}
\eta^{12}=\eta^{21}=1 \qquad \text{and} \qquad \eta^{44}=-2~. \label{etadef2}
\end{align}

The full six-dimensional geometry, constrained by supersymmetry, may be written as a $(1+1)$-dimensional Lorentzian fiber, parametrized by the light cone coordinates $(u,v)$ (see (\ref{uvDef})), over a four dimensional hyper-K\"ahler base $ds_4^2(\mathcal{B})$ as: 
\begin{align}
ds_6^2 &=   -\frac{2}{\sqrt{\mathcal{P}}} \, (dv+\beta) \big(du +  \omega + \tfrac{1}{2}\, \mathcal{F} \, (dv+\beta)\big) 
+  \sqrt{\mathcal{P}} \, ds_4^2(\mathcal{B})~. \label{ds6}
\end{align}
The metric data consists of the base $\mathcal{B}$, the functions $(\mathcal{P},\mathcal{F})$ and one forms $(\beta,\omega)$. The one forms must have legs only on the base $\mathcal{B}$, while the complete data can have functional dependence on all coordinates except for $u$. This metric data is fixed by solving the BPS equations, which are written in terms of a set of three functions $Z_{\hI}$ and three two forms $\Theta^{\hI}$. In terms of this data, the three-form fields encoding the multiplets read:
\begin{align}
G^{\hat{I}} =  d \left[ -\frac{1}{2}\frac{\eta^{\hat{I}\hat{J}}Z_{\hat{J}}}{\mathcal{P}}(du+\omega)\wedge (dv+\beta) \right] + \frac{1}{2}\eta^{\hat{I}\hat{J}}*_{4}DZ_{\hat{J}}+\frac{1}{2}(dv+\beta)\wedge \Theta^{\hat{I}} \label{Gdef}
\end{align}
where:
\begin{align}
\mathcal{P} = \frac{1}{2}\eta^{\hat{I}\hat{J}}Z_{\hat{I}}Z_{\hat{J}}= Z_{1}Z_{2}-(Z_{4})^{2}~,
\end{align}
$D$ is defined to act on forms $\Phi$ by:
\begin{align}
D\Phi=d_{4}\Phi-\beta\wedge \dot{\Phi}
\end{align}
where overhead dots denote $\partial_{v}$ derivatives, $(d_{4},*_{4})$ are the exterior derivative and Hodge star with respect to the four-dimensional hyper-K\"ahler base $\mathcal{B}$, and their non-subscript and hatted counterparts refer to the full six-dimensional geometry (\ref{ds6}). These three forms satisfy the twisted self duality constraint:\footnote{Note that this corresponds to choosing $\varepsilon=+1$ in (\ref{eq:6Dselfduality}). Also note that our Hodge dual conventions (as given in (\ref{eq:6Dselfduality})) imply that there should indeed be two anti-self-dual tensors and one self-dual tensor for the superstrata.}
\begin{align}
\hat{*} G^{\hat{I}} = \tensor{M}{^{\hat{I}}_{\hat{J}}}G^{\hat{J}} \qquad \text{where} \qquad M_{\hat{I}\hat{J}} =  \frac{Z_{\hat{I}}Z_{\hat{J}}}{\mathcal{P}}-\eta_{\hat{I}\hat{J}}~. \label{M1mnDef}
\end{align}

Solving the BPS equations takes the layered form:
\begin{itemize}
\item Fix a hyper-K\"ahler base $ds_4^2(\mathcal{B})$ and choose a $\beta$ satisfying: 
\begin{align}
d\beta = *_{4}d\beta ~.\label{BPSlayer0}
\end{align}
\item Find a set $(Z_{\hat{I}},\Theta^{\hat{I}})$ that solve the ``first layer"\footnote{Significant process was made in solving this layer in general in \cite{Tyukov:2018ypq}. For any harmonic functions $\Phi_{\hI}$ on $\mathcal{B}$, one can derive from them and a complex structure a self dual two forms $\Theta^{\hI}$.  If it is known what modulus of $\mathcal{B}$ these two form control as a K\"{a}hler deformation, then the $Z_{\hI}$ which solve the first BPS layer together with these $\Theta^{\hI}$ can be found directly from the $\Phi_{\hI}$.}:
\begin{align}
*_{4}D\dot{Z}_{\hat{I}} = \eta_{\hat{I}\hat{J}}D\Theta^{\hat{J}} ~,\qquad D*_{4}DZ_{\hat{I}}=-\eta_{\hat{I}\hat{J}} \Theta^{\hat{J}}\wedge d\beta ~, \qquad \Theta^{\hat{I}} = *_{4} \Theta^{\hat{I}}~. \label{BPSlayer1}
\end{align} 
\item Find $(\mathcal{F},\omega)$ that solve the ``second layer:"
\begin{align}
(1+*_{4})D\omega +\mathcal{F}\, d\beta &= Z_{\hat{I}}\Theta^{\hat{I}} ~, \label{BPSlayer2a}\\
*_{4}D*_{4} \left(\dot{\omega}- \frac{1}{2}D\mathcal{F} \right) &= \frac{1}{4}\eta_{\hat{I}\hat{J}}\left[ 4\ddot{Z}^{\hat{I}}Z^{\hat{J}}+2\dot{Z}^{\hat{I}}\dot{Z}^{\hat{J}} - *_{4}\left( \Theta^{\hat{I}}\wedge \Theta^{\hat{J}} \right)\right]~. \label{BPSlayer2b}
\end{align}
\end{itemize}

\subsection{The solution}
The standard hyper-K\"ahler base used in the construction of six-dimensional superstrata is flat $\mathbb{R}^{4}$, which is most conveniently written in spherical bipolar coordinates $(r,\theta,\varphi_{1},\varphi_{2})$, with metric:
\begin{equation}
ds_4^2(\mathcal{B}) ~=~ \Sigma \, \left(\frac{d r^2}{r^2+a^2}+ d\theta^2\right)+(r^2+a^2)\sin^2\theta\,d\varphi_1^2+r^2 \cos^2\theta\,d\varphi_2^2\, .
 \label{ds4flat}
\end{equation}
where $a$ is a positive constant and
 \begin{equation}
\Sigma~\equiv~  r^2+a^2 \cos^2\theta     \,.
 \label{Sigdefn}
\end{equation}

In terms of the complex coordinates: 
\begin{equation}
\chi ~\equiv~\frac{a}{\sqrt{r^2+ a^2}} \, \sin \theta \, e^{i \varphi_1} \,, \qquad \mu ~=~\cot \theta \, e^{i \big (\frac{\sqrt{2} v}{R_y} - \varphi_1- \varphi_2\big)} \,, \qquad \xi ~\equiv~\frac{r}{\sqrt{r^2+ a^2}} \, e^{i \frac{\sqrt{2} v}{R_y} }\, , \label{cplxcoorddefn}
\end{equation}
the $(1,m,n)$ multi-mode solution can be written in terms of the basic function:
\begin{align}
F(\chi,\mu,\xi)= \chi F_{0}(\xi)+\chi \mu F_{1}(\xi)~, 
\end{align}
where $F_{0,1}$ are holomorphic functions of $\xi$, with expansions in terms of the real coefficients $(c_{n},d_{n})$:
\begin{align}
F_{0}=\sum_{n=1}^{\infty}b_{n}\xi^{n} \qquad \text{and} \qquad F_{1}=\sum_{n=1}^{\infty}d_{n}\xi^{n}~. 
\end{align}

We define the auxiliary data:
\begin{align}
A= \chi\mu\left(1+\xi\partial_{\xi} \right)F_{1} \qquad \text{and} \qquad B=\chi\xi\partial_{\xi}F_{0}   \, , \label{ABdef}
\end{align}
and self dual forms 
\begin{align}
\Omega_{y} ~=~  \frac{1}{\sqrt{2}} \left(-\Omega^{(2)}+i r\sin\theta \, \Omega^{(1)} \right)\, ,  \qquad
\Omega_{z} ~=~  \frac{1}{\sqrt{2}}\left(\Omega^{(3)}+i \left(r\sin\theta - \frac{\Sigma}{r\sin\theta} \right)  \Omega^{(1)} \right)\, . \label{SDyz}
\end{align}
where 
\begin{equation}
\label{selfdualbasis}
\begin{aligned}
\Omega^{(1)} &~\equiv~ \frac{dr\wedge d\theta}{(r^2+a^2)\cos\theta} + \frac{r\sin\theta}{\Sigma} d\varphi_1\wedge d\varphi_2\,,\\
\Omega^{(2)} &~\equiv~  \frac{r}{r^2+a^2} dr\wedge d\varphi_2 + \tan\theta\, d\theta\wedge d\varphi_1\,,\\
 \Omega^{(3)} &~\equiv~ \frac{dr\wedge d\varphi_1}{r} - \cot\theta\, d\theta\wedge d\varphi_2\,.
\end{aligned}
\end{equation}

To solve the BPS equations, first, one fixes:
\begin{align}
\beta = \frac{a^{2}R_{y}}{\sqrt{2}} \left(\sin^{2}\theta \, d\varphi_{1} -\cos^{2} \theta \, d\varphi_{2} \right)\,, \label{beta}
\end{align}
then the solution to the first BPS layer is given by the data:
\begin{align}
\begin{split}
Z_{1} &= \frac{Q_{1}}{\Sigma} + \frac{R_y^{2}}{4Q_{5}\Sigma} \left( F^{2}+\bar{F}^{2}\right) \,, \\
Z_{2}&= \frac{Q_{5}}{\Sigma} \,, \\
Z_{4} &= \frac{R_y}{2\Sigma} \left(F+\bar{F} \right) \,, 
\end{split}
~~
\begin{split}
 \Theta^1 &= 0 \,,\\
\Theta^2 &=   \frac{R_y}{Q_{5}}\,F  \left(A\, \Omega_{y} + B\, \Omega_{z}  \right)\,+\,c.c. \,, \\
\Theta^4 &= -2\left( A\, \Omega_{y} + B\, \Omega_{z} \right)+\,c.c. \,.
\end{split}
\label{1stLayerGenHolo}
\end{align}

The solution to the second BPS layer can then be written in the form:
\begin{align}
\mathcal{F}&= \mathcal{F}^{(p)}+ c^{2}\mathcal{F}^{(c)}\\
\omega &=\frac{4}{\sin 2\theta}\omega_{\mu}^{(p)} \, d\theta +2\left(\omega_{\chi}^{(0)}+\omega_{\chi}^{(p)}+c^{2}\,\omega_{\chi}^{(c)} \right)\, d\varphi_{1} +2\left(\omega_{\delta}^{(0)}+\omega_{\delta}^{(p)} \right) \, d\varphi_{2}~, \label{FomegaAnsatz2}
\end{align}
where $c$ is a constant.\footnote{Note that we introduce this constant of integration as $c^2$, whereas $c$ (i.e. unsquared) was used in \cite{Heidmann:2019xrd}. As we will see in (\ref{cDef2}), $c^2$ is naturally a positive number.} The ``round supertube" part is given by:
\begin{align}
\omega^{(0)}_{\chi}= \frac{\omega^{(0)}_{\delta}}{\abs{\mu}^{2}}  = \frac{R_{y}\abs{\chi}^{2}}{2\sqrt{2}(1-\abs{\chi}^{2})}~. 
\end{align}
The homogeneous part is given by:
\begin{align}
\mathcal{F}^{(c)}&= - \frac{1}{a^{2}} \qquad \text{and} \qquad \omega^{(c)}_{\chi} =  \frac{R_{y} \abs{\chi}^{2}}{2\sqrt{2}a^{2}(1-\abs{\chi}^{2})}~.
\end{align}
Finally, the solution is completed by adding the particular part:
\begin{align}
\mathcal{F}^{(p)}&= \frac{1}{a^{2}} \left(\abs{F_{0}}^{2}+\abs{\xi}^{2} \abs{F_{1}}^{2}  \right) \,, 
&\omega_{\chi}^{(p)}&= - \frac{R_{y} }{4\sqrt{2}a^{2}(1-\abs{\chi}^{2})}\left( \bar{\chi}\bar{F}_{0}F+\chi F_{0}\bar{F}\right) \,,\\
\omega_{\mu}^{(p)} &= - \frac{iR_{y}\abs{\chi}^{2}}{4\sqrt{2}a^{2}} \left( \mu \bar{F}_{0}F_{1}-\bar{\mu}F_{0}\bar{F}_{1} \right) \,,
&\omega_{\delta}^{(p)}&= \frac{R_{y} \abs{\xi}^{2}}{4\sqrt{2}a^{2}(1-\abs{\chi}^{2})} \left( \chi \mu F_{1}\bar{F}+\bar{\chi}\bar{\mu}\bar{F}_{1}F\right) \,.
\end{align}

\subsection{Tuning the asymptotic geometry}
\label{SubSect:Asypmtotic Geom}
Setting $F_{0}=F_1=0$ in the $(1,m,n)$ solution of the previous section gives the ``round supertube" solution, which is globally $AdS_{3}\times S^{3}$. This solution has
\begin{align}
\cF = 0 \qquad \text{and} \qquad \omega = \omega_{0} =  \frac{a^{2}R_{y}}{\sqrt{2}} \left(\sin^{2}\theta \, d\varphi_{1} +\cos^{2} \theta \, d\varphi_{2} \right) \,. \label{Supertube}
\end{align}

To ensure the $(1,m,n)$ solutions have the same asymptotics, one must arrange for $(\cF,\omega)$ to have at most $\mathcal{O}(r^{-2})$ corrections to the round supertube solution (\ref{Supertube}). We achieve this by first defining:
\begin{align}
F_{0}^{(\infty)}(v) \equiv \lim_{\abs{\xi}\to\infty}F_{0}(\xi)=\lim_{r\to\infty}F_{0}(\xi) \qquad \text{and} \qquad F_{1}^{(\infty)}(v) \equiv \lim_{\abs{\xi}\to\infty}F_{1}(\xi)=\lim_{r\to\infty}F_{1}(\xi) ~, \label{AsymtoticFs}
\end{align}
and then fixing $c^{2}$ by:
\begin{align}
c^{2}\equiv \frac{1}{\sqrt{2}\pi R} \int_{0} dv \, \left( \abs{F_{0}^{(\infty)}(v)}^{2}+\abs{F_{1}^{(\infty)}(v)}^{2}\right)=b^{2}+d^{2}~, \label{cDef2}
\end{align}
where we have implicitly defined
\begin{align}
b^{2}=\sum_{n=1}^{\infty}b_{n}^{2} \qquad \text{and} \qquad d^{2}=\sum_{n=1}^{\infty}d_{n}^{2}~. \label{c01Def}
\end{align}
Then, we can use a gauge transformation which leaves the six-dimensional BPS equations (\ref{BPSlayer0})-(\ref{BPSlayer2b}) and metric (\ref{ds6}) invariant:
\begin{align}
u \to u+f(v,r,\theta,\varphi_{1},\varphi_{2}) \qquad \iff \qquad \omega \to \omega - d_{4}f+ \dot{f}\beta \,, \qquad \cF\to \cF-2\dot{f}\,,
\end{align}
with gauge parameter chosen as:
\begin{align}\label{eq:fdef}
f(v)\equiv \frac{1}{2a^{2}}\int_{0}^{v} dv' \, \left( \abs{F_{0}^{(\infty)}(v')}^{2}+\abs{F_{1}^{(\infty)}(v')}^{2}-c^{2}\right)\,,
\end{align}
which brings the $(\cF,\omega)$ for the $(1,m,n)$ family to the form
\begin{align}
\cF &= -\frac{1}{r^{2}} \left(d^{2}+ \sum_{n=1}n(b_{n}^{2}+d_{n}^{2}) + {\it oscillating \ terms} \right) + \mathcal{O}(r^{-4})  \,, \label{FAsym} \\
\omega &= \omega_{0} + \frac{R_{y}d^{2}}{\sqrt{2}r^{2}} \left(   \sin^{2}\theta\, d\varphi_{1} +\cos^{2}\theta \, d\varphi_{2}+ {\it oscillating \ terms}\right) + \mathcal{O}(r^{-4})\,. \label{omegaAsym}
\end{align}
This gauge-transformed geometry is now asymptotically the same as the round supertube, i.e. $AdS_{3}\times S^{3}$. Note that the relation (\ref{cDef2}) is crucial to make the gauge parameter (\ref{eq:fdef}) a well-defined, periodic function of $v$; without imposing (\ref{cDef2}) it is not possible to retrieve the correct asymptotics.

\subsection{Regularity and CTC analysis}
\label{SubSect:Regularity}
There are four distinct ways in which the six-dimensional metric (\ref{ds6}), for the $(1,m,n)$ solution of the previous section, which also takes the form (\ref{eq:6Dmetricansatz}), may fail to be regular:
\begin{itemize}
\item The metric is singular where the data $(\beta,\omega,\mathcal{F})$ are singular, at the locus:
\begin{align}
\Sigma = 0~. \label{SigmaLocus}
\end{align}
\item The warp factors $(\Delta^{-1}\det m_{AB})^{\pm 1/2}$ are singular. 
\item The sphere deformations $\tensor{\widetilde{A}}{_{\mu}^{AB}}$ of (\ref{Aform}) are singular. 
\item The $ds_{3}^{3}$ metric (\ref{ds3})-(\ref{ds3Data}) possesses a conical singularity at $r=0$, where the $y$-circle pinches off. 
\item $ds_{3}^{3}$ possesses closed time-like curves (CTCs).
\end{itemize}

Upon expanding and analyzing the metric along the locus (\ref{SigmaLocus}), the only potentially singular part was found to be the $d\varphi_{1}^{2}$ coefficient. Setting
\begin{align}
r=a \epsilon \qquad \text{and} \qquad \theta=\frac{\pi}{2}-\epsilon~,
\end{align}
and expanding in powers of $\epsilon$, this term reads:
\begin{align}
\frac{1}{\epsilon^{2}}\left(\frac{2 Q_{1}Q_{5}-(2a^{2}+c^{2})R_y^{2}}{2\sqrt{4Q_{1}Q_{5}-2\abs{F_{0}}^{2}R_y^{2}}}\right)\, d\varphi_{1}^{2} + O(\epsilon^0)~.
\end{align}
To remove this singularity one must tune: 
\begin{align}
\frac{Q_{1}Q_{5}}{R_{y}^{2}}=\frac{1}{g_{0}^{4}R_{y}^{2}}=a^{2}+\frac{c^{2}}{2}~. \label{RegCond}
\end{align}

Now consider 
\begin{align}
\det m_{AB} &=\left( 1-S_{A}S_{A}\right)^{2} =\left(1-  \abs{z_{1}}^{2} -\abs{z_{2}}^{2} \right)^{2} \,,
\end{align}
where 
\begin{align}
z_{1}= S_{1}+iS_{2} \qquad \text{and} \qquad z_{2}=S_{3}+iS_{4}\,.
\end{align}
Since $m^{AB}=(m_{AB})^{-1}$ appears in the kinetic term of the three-dimensional Lagrangian (\ref{action-final}), its solutions must necessarily bound $\det m_{AB}$ away from zero\footnote{This argument is plausible, rather than providing a strict proof. The scalar action would become infinite wherever $\det m_{AB}=0$, so the minimization procedure should ensure solutions avoid this condition.}. Since $\lim_{r\to\infty}S_{A}=0$ we conclude that:
\begin{align}
0<(\det m_{AB})^{1/2} = 1-  \abs{z_{1}}^{2} -\abs{z_{2}}^{2}  \,, \label{detmPositive}
\end{align}
for all solutions of the action (\ref{action-final}), which includes the $(1,m,n)$ solution. Now one can also calculate that 
\begin{align}
\Delta = 1 - \abs{\tilde{z}_{1}+\tilde{z}_{2}}^{2}\,,
\end{align}
where
\begin{align}
\tilde{z}_{1} = (S_{1}-iS_{2})\sin\theta \, e^{-i\varphi_{1}} \qquad \text{and} \qquad \tilde{z}_{2} = -(S_{3}+iS_{4})\cos\theta \, e^{i\varphi_{2}}\,.
\end{align}
The triangle inequality and (\ref{detmPositive}) then imply that 
\begin{align}
0<\Delta\,. 
\end{align}
Hence the warp factors $(\Delta^{-1}\det m_{AB})^{\pm 1/2}$ are regular. 

In passing, looking at the form of $\Omega^{2}$ in (\ref{ds3Data}), it also follows that:
\begin{align}
0<\Omega^{2}\,. \label{Omega2Positive}
\end{align}
Hence the sphere deformations $\tensor{\widetilde{A}}{_{\mu}^{AB}}$ of (\ref{Aform}) are clearly regular by inspection. 

Setting $\rho=r/a$, the three dimensional metric can be written as:
\begin{align}
ds_{3}^{2} &= \frac{1}{g_{0}^{2}}\left[\frac{d\rho^{2}}{1+\rho^{2}}-g_{0}^{8}a^{4}R_{y}^{2}(1+\rho^{2})dt^{2} +\frac{\rho^{2}}{R_{y}^{2}}\left(dy +(1-g_{0}^{4}a^{2}R_{y}^{2})dt \right)^{2}  \right] \notag \\
& \qquad \qquad \qquad \qquad \qquad - \frac{g_{0}^{2}}{2}\left( \abs{F_{0}}^{2}+\abs{F_{1}}^{2} \right) \left[ \rho^{2}(dt+dy)^{2}+\frac{R_{y}^{2}\,d\rho^{2}}{(1+\rho^{2})^{2}}  \right]\,. \label{ds3SupertubePert}
\end{align}
This form of the metric makes it clear that there is no conical singularity when the $y$-circle pinches off at $\rho=0$. 

Proving there are no CTCs in (\ref{ds3SupertubePert}), when suitably regularized by (\ref{RegCond}), and tuned to be asymptotically $AdS_{3}$ by (\ref{cDef2}), is a delicate business. A proof for the $(1,0,n)$ and $(1,1,n)$ families appears in \cite{Heidmann:2019xrd}. It relies on properties of $\abs{F_{0,1}}^{2}$ following from the analyticity of $F_{0,1}$, which do not easily generalize to the sum  $\abs{F_{0}}^{2}+\abs{F_{1}}^{2}$, as is required for the $(1,m,n)$ family. Although we do not have a proof, we expect $ds_{3}^{2}$ to be free of CTCs when tuned with (\ref{cDef2}) and (\ref{RegCond}) for the $(1,m,n)$ family. This expectation is based on examining many examples with explicit expansions of $F_{0,1}$, as well as the fact that the holographic duals should be well defined CFT states. Intuitively, one can think of the second line of (\ref{ds3SupertubePert}) as a ``perturbation" of a regular $AdS_{3}$ seed. Upon fixing the CFT charges $Q_{1,5}$, the magnitude of $\abs{F_{0,1}}^{2}$ are restricted by (\ref{cDef2}) and (\ref{RegCond}), and so the negative contribution coming from the ``perturbation" is sufficiently controlled so as to avoid CTCs.

\subsection{Conserved charges}
A detailed analysis of computing the conserved charges for the six-dimensional superstrata appears in \cite{Heidmann:2019xrd}, where the explicit calculations for the $(1,0,n)$ and $(1,1,n)$ multi-mode families are also given. Here we give a short summary of the procedure, which are closely analogous to the individual $(1,0,n)$ and $(1,1,n)$ family analysis, and the result. The analysis requires $c^{2}$ to be tuned as in (\ref{cDef2}), so that the geometry is asymptotic to $AdS_{3}\times S^{3}$.

The D1-D5-P system possesses five conserved charges: the net charge of each type of brane $Q_{1,5}$, the momentum in the common D1-D5 direction $Q_{P}$, and the two angular momenta $J_{L,R}$. The brane charges $Q_{1,5}$ can be simply read off from the $r^{-2}$ coefficient of $Z_{1,2}$ in (\ref{1stLayerGenHolo}), when expanded about $r\to\infty$.

Using the solution as presented in the gauge of (\ref{FAsym})-(\ref{omegaAsym}), with $c^{2}$ fixed by (\ref{cDef2}), the remaining charges can be read off from the expansions:
\begin{equation}
\beta_1+\beta_2 +\omega_1 + \omega_2~=~ \frac{ \sqrt{2}}{r^2}\, \big[ \, (J_R - J_L \cos 2\theta )  ~+~ {\it oscillating \ terms}\, \big] ~+~ \mathcal{O}(r^{-4}) 
\label{asmpmoms}
\end{equation} 
where $ \beta_1, \omega_1$ and $ \beta_2, \omega_2$ are the components of $\beta$ and $\omega$ along $d\varphi_1$ and $d\varphi_2$ respectively, and
\begin{equation}
\mathcal{F}  ~=~ - \frac{1}{r^2} \,\big(2\, Q_P  ~+~{\it oscillating \ terms}   \big) ~+~ \mathcal{O}(r^{-4})   \,.
\label{Fexp}
\end{equation}
For the $(1,m,n)$ multi-mode solution this procedure gives (using (\ref{c01Def})):
\begin{align}
J_{L}=\frac{a^{2}R_{y}}{2} \,, \qquad  \qquad J_{R}  = \frac{R_{y}}{2}\left(a^{2}+d^{2} \right) \,, \qquad Q_{P} = \frac{1}{2}\left[ d^{2}+\sum_{n=1}^{\infty} n \left(b_{n}^{2} + d_{n}^{2}  \right) \right]\,.
\end{align}

\subsection{Uplifting the six-dimensional scalars to ten dimensions}
The parametrization of the six-dimensional scalars $\varphi,X$ that we used in the uplift formulae in Section \ref{sec:full6Duplift} is convenient from the point of view of the six-dimensional reduction, as $\varphi$, resp. $X$, only depend on $m_{AB}$, resp. $\chi_A$. However, for the D1-D5 system, they require some rearranging to consider the uplift to ten dimensions; in particular, the ten-dimensional dilaton is not given by $\varphi$. With the following rearrangement of the scalars, introducing an arbitrary constant $\tilde{Z}_2$:
\be e^{2\phi^{\text{(10D)}}} = \frac{e^{\sqrt{2} \varphi } \left(2e^{-\sqrt{2} \varphi }+X^2\right)^2}{4\tilde{Z}_2^2}, \qquad C_{(0)}^{\text{(10D)}}=-\sqrt{2}\tilde{Z}_2\frac{  X }{2e^{-\sqrt{2} \varphi }+X^2},\ee
the six-dimensional scalar kinetic action becomes:
\be \mathcal{L}_{6D,\text{scal.kin}} = -\frac12 (\partial\varphi)^2 -\frac12 e^{\sqrt{2}\varphi}(\partial X)^2 = -(\partial \phi^{\text{(10D)}})^2 - e^{2\phi^{\text{(10D)}}}(\partial C_{(0)}^{\text{(10D)}})^2.\ee
Then, for the superstrata uplift to ten dimensions (using the conventions of, for example, Appendix B of \cite{deLange:2015gca}), $\phi^{\text{(10D)}}$ can be identified with the ten-dimensional dilaton after uplifting on a $\IT^4$ while $C_{(0)}^{\text{(10D)}}$ is the ten-dimensional axion. Setting $\tilde{Z}_2 = Z_2(g_0^2\Sigma)$ then gives the traditional form for these scalars:
\be e^{2\phi^{\text{(10D)}}} = \frac{Z_1^2}{\mathcal{P}}, \qquad C_{(0)}^{\text{(10D)}} = \frac{Z_4}{Z_1},\ee
whereas the six-dimensional fields used in Section \ref{sec:full6Duplift} were simply: 
\be e^{-\sqrt{2}\varphi} = (g_0^2\Sigma)^2\mathcal{P}, \qquad X = -\sqrt{2}(g_0^2\Sigma) Z_4.\ee

\section{Three-dimensional equations of motion}
\label{App:3Deom}
Here, we will give for reference the equations of motion following from the three-dimensional action (\ref{action-final}).

The scalar equations of motion are:
\begin{align}
0&= \cD_\mu \cD^\mu \chi_A
 - \cD^\mu\chi_B \left(\varepsilon\, \epsilon\ind{_\mu^{\nu\rho}}F^{BC}_{\nu\rho}m_{CA} 
 +m^{BC}  \cD_\mu m_{CA}\right)\\
\nonumber & -(\det m^{-1})\left( 8\varepsilon\, \alpha \gamma_0 m_{AB}\chi_B +4\gamma_0^2  m_{AB}m_{BC}\chi_C + 2\gamma_0^2m_{AB}\chi_B (\chi_C\chi_C) \right),\\
0 &= \cD_\mu \cD^\mu m_{AB}
+ \cD_\mu \chi_A \cD^\mu \chi_B
-m^{CD}\cD^\mu m_{AC} \cD_\mu m_{BD}
-2 F^{CD}_{\mu\nu} F^{EF\, \mu\nu} m_{AC}m_{BE}m_{DF}\\
\nonumber &+ (\det m^{-1})\left( \left[16\alpha^2+8\gamma_0^2m_{CD}m_{CD}-4\gamma_0^2m_{CC}m_{DD}\right]m_{AB} \right.\\
\nonumber & \left. +8\gamma_0^2m_{CC}m_{AD}m_{BD} -16\gamma_0^2m_{CD}m_{AC}m_{BD} \right.\\
\nonumber & \left.+ 8\varepsilon\, \gamma_0\alpha m_{AB} \chi_C\chi_C + 4\gamma_0^2(m_{CD}m_{AB}-m_{AC}m_{BD})\chi_C\chi_D +\gamma_0^2m_{AB}(\chi_C\chi_C)^2 \right),
\end{align}
the Einstein equations are:
\begin{align}
 R_{\mu\nu} - \frac12 R g_{\mu\nu} &= 2 g_{\mu\nu}\left( -~\coeff{1}{16}\,  {\rm Tr}\big[ \big( \cD_\rho  m \big) m^{-1}\, \big( \cD^\rho m \big) m^{-1}\,  \big]   \right.\\
 \nonumber & \left. -\coeff{1}{8}\, m^{AB} \,(\cD_\rho\chi_A)\,(\cD^\rho \chi_B) ~-~V - \coeff{1}{8} m_{AC} \,m_{BD}\, F_{\rho \sigma}^{AB}\, F^{\rho \sigma}{}^{CD} \right)\\
\nonumber  & + \frac14 m^{AC}m^{BD}\cD_\mu m_{AB}\cD_\nu m_{CD} + \frac12 m^{AB}\cD_\mu \chi_A \cD_\nu \chi_B + m_{AC}m_{BD} F\ind{^{AB}_{\mu}^{\rho}}F^{CD}_{\nu\rho},
\end{align}
and finally the gauge field equations of motion are:
\begin{align}
 0&= m_{AC}m_{BD}\cD_\nu F\ind{^{CD}_\mu^\nu} + \left( F^{CD}_{\mu\nu} m_{BD} \cD^\nu m_{AC} - F^{CD}_{\mu\nu} m_{AD} \cD^\nu m_{BC}  \right)\\
 \nonumber & +\gamma_0\epsilon_{ABCD}m^{DE}( \cD_\mu m_{CE} + \chi_C \cD_\mu \chi_E) - \alpha\epsilon_{ABCD} \epsilon\ind{_\mu^{\nu\rho}} F^{CD}_{\nu\rho}\\
  \nonumber &  +\frac12\varepsilon\left( -\epsilon_{\mu\nu\rho} \cD^\nu\chi_A \cD^\rho \chi_B +\frac12 \epsilon_{\mu\nu\rho} (\chi_B\cD^\nu \cD^\rho\chi_A-\chi_A\cD^\nu \cD^\rho\chi_B)    - \gamma_0\epsilon\ind{_\mu^{\nu\rho}}\epsilon_{ABCD}F^{ED}_{\nu\rho}\chi_E\chi_C \right) . 
\end{align}

\newpage

\begin{adjustwidth}{-1mm}{-1mm} 

\bibliographystyle{utphys}      

\bibliography{microstates}       

\end{adjustwidth}

\end{document}